\documentclass[linenumbers, astrosymb, tighten, twocolumn, fleqn]{cas-dc}

\usepackage{amsmath}
\usepackage{lineno}
\usepackage{mathtools}
\usepackage{url}
\usepackage{gensymb}
\usepackage{subcaption}
\usepackage{setspace}

\usepackage{cancel}
\usepackage{xspace}
\usepackage{xfrac}

\newcommand{\de}{{\rm d}}
\newcommand{\DIAD}{\texttt{DIAD}\xspace}
\newcommand{\RAD}{\texttt{RADMC-3D}\xspace}
\newcommand{\RADp}{\texttt{RAD+}\xspace}

\usepackage[range-phrase=-, range-units=single, list-final-separator=\text{, and }]{siunitx}

\usepackage[authoryear]{natbib}

\shorttitle{Circumplanetary Disk Structure}
\shortauthors{Taylor, Adams,  \& Calvet}

\begin{document}


\title[mode = title]{The Two-Dimensional Structure of Circumplanetary Disks \\
and their Radiative Signatures}

\author[1]{Aster G. Taylor}[orcid=0000-0002-0140-4475]
\cormark[1]
\fnmark[1]
\ead{agtaylor@umich.edu}
\affiliation[1]{organization={Department of Astronomy, University of Michigan},
    city={Ann Arbor},
    postcode={48109}, 
    state={MI},
    country={USA}}

\author[2,1]{Fred C. Adams}[orcid=0000-0002-8167-1767]
\affiliation[2]{organization={Department of Physics, University of Michigan},
    city={Ann Arbor},
    postcode={48109}, 
    state={MI},
    country={USA}}

\cortext[cor1]{Corresponding author}
\fntext[fn1]{Fannie and John Hertz Foundation Fellow}

\author[1]{Nuria Calvet}[orcid=0000-0002-3950-5386]

\begin{abstract}
During their formative stages, giant planets are fed by infalling material sourced from the background circumstellar disk. Due to conservation of angular momentum, the incoming gas and dust collects into a circumplanetary disk that processes the material before it reaches the central planet itself. This work investigates the complex vertical structure of these circumplanetary disks and calculates their radiative signatures. A self-consistent numerical model of the temperature and density structure of the circumplanetary environment reveals that circumplanetary disks are thick and hot, with aspect ratios $H/R\sim0.1-0.25$ and temperatures approaching that of the central planet. The disk geometry has a significant impact on the  radiative signatures, allowing future observations to determine critical system parameters. The resulting disks are gravitationally stable and viscosity is sufficient to drive the necessary disk accretion. However, sufficiently rapid mass accretion can trigger a thermal instability, which sets an upper limit on the mass accretion rate. This  paper shows how the radiative signatures depend on the properties of the planetary system and discuss how the system parameters can be constrained by future observations. 
\end{abstract}

\maketitle


\section{Introduction}\label{sec:intro}

Thirty years after the discovery of the first exoplanet around a main-sequence star \citep{Mayor1995}, nearly \num{6000} exoplanets have been discovered. While this sample has contributed to an explosion of scientific research, how these systems form is still not well-understood. Observational constraints on the physics and structure of planet accretion, including the role played by circumplanetary disks, are essential to improving our understanding of giant planet formation. Although only a few circumplanetary disks have been detected to date \citep{Benisty2021, Christiaens2024, Cugno2024, Fasano2025}, the advent of \textit{JWST} and the forthcoming Extremely Large Telescope (ELT) are expected to detect and characterize more of these systems. Accurate models of forming planets and their surrounding environment are needed to analyze and interpret these observations. Toward this end, this paper investigates the structure and radiative signatures of circumplanetary disks during the late stages of giant planet formation. 

Most giant planets are expected to form through the core accretion process, which can be divided into three stages \citep{Pollack1996}. During the first stage, solid material collects into a $\sim\qty{10}{M_\oplus}$ core. The object then begins to accrete gaseous material and forms a hydrostatically-supported envelope that extends out to the Hill radius. The envelope slowly cools and contracts, so this stage represents a bottleneck in the planet formation process. After the protoplanet reaches a mass $M_p\sim\qty{20}{M_\oplus}$, the envelope is no longer fully supported by pressure and contracts more rapidly. The forming planet then enters a stage of runaway mass accretion. The planet gains most of its mass during this third stage.  

Circumplanetary disks are expected to form during this final stage (see, e.g., \citealt{Lubow1999, Canup2002}). Through a process analogous to the formation of circumstellar disks, angular momentum conservation does not allow infalling material to accrete directly onto the planet. Instead, the material will fall to the equatorial plane, dissipate its vertical velocity in a strong shock, and form a nearly Keplerian circumplanetary disk. Most of the planetary mass is processed through the disk, rather than falling directly onto the surface of the planet. Although most of the mass that falls to the disk subsequently moves inward and onto the planet, some fraction is transferred outward to conserve angular momentum. 

Although many hydrodynamic simulations of giant planet formation predict the existence of circumplanetary disks (e.g., \citealt{Szulagyi2016, Lambrechts2019}), the structure of these disks varies widely. For example, some studies (e.g., \citealt{Ayliffe2009}) show that incoming material is accreted along equatorial directions, while others (e.g., \citealt{Lambrechts2017}) indicate that material primarily enters through the poles of the system. These qualitative disagreements suggest that simulations have not yet converged on the expected structure or the formation processes of circumplanetary disks. 

Previous work has addressed this issue by analytically determining the structure of the circumplanetary system \citep{Adams2022, Taylor2024}. That work calculated how the envelope density, the surface density of the disk, and the luminosities of the planet and the disk depend on relevant system parameters, including the accretion geometry. More recent work calculated the temperature distribution of the envelopes and presented images and spectral energy distributions (SEDs) of the forming systems \citep{Taylor2025}. 

This previous work accurately modeled the infalling circumplanetary envelope, but relied on several simplifying assumptions concerning the circumplanetary disk itself. Specifically, these previous calculations assumed that (i) the disk is confined to the equatorial plane, (ii) the disk temperature distribution has a power-law form $r^{-p}$ where $p=1/2-3/4$, and (iii) the disk emits as an optically-thick blackbody at all points on its surface. The present work generalizes the previous treatment by constructing self-consistent two-dimensional numerical models of circumplanetary disks and by calculating their corresponding radiative signatures. These models not only provide an improved description of the structure of circumplanetary disks, but also enable future observations to better constrain the properties and relevant physics of accreting protoplanetary systems. 

This paper is structured as follows. In Section~\ref{sec:models} we discuss our numerical model and briefly review our previous semianalytic model. Section~\ref{sec:diskstruc} calculates the structure of the circumplanetary disks predicted by the numerical model and discusses the implications of this structure. The radiative signatures that result from these disks and how these signatures vary with system properties are explored in Section \ref{sec:disksig}. 
The paper concludes in Section~\ref{sec:disc} with a summary of our results and a discussion of their implications. 

\section{Numerical Model}\label{sec:models}

This section presents the numerical model used to calculate the circumplanetary disk structure and its corresponding SEDs. 
Two numerical packages are used to self-consistently calculate the disk structure --- the D'Alessio Irradiated Accretion Disk model \citep[\DIAD,][]{DAlessio1998, DAlessio1999, DAlessio2001, DAlessio2006} and the radiative transfer Monte Carlo code \RAD \citep{Dullemond2012}. 

The temperature and density structure of the disk are determined through an iterative process, with the \DIAD code providing an initial estimate for the disk density distribution. This initial density estimate is interpolated into a two-dimensional {azimuthally symmetric} spherical geometry, with \num{250} (logarithmically spaced) points in the $r$ direction and \num{150} points in the $\theta$ direction. There are \num{50} additional points from {a factor of} \numrange{0.9}{1.1} {inner wall radii} to ensure that the wall is well-resolved.\footnote{{Models with increased resolution exhibit nearly identical structures and radiative signatures, indicating that this model has converged at this resolution.}} This density structure is then loaded into \RAD. Since the disk is symmetric across the midplane, $\theta$ ranges from $0$ to $\pi/2$. 

We then use \RAD to calculate the temperature throughout the two-dimensional region characterized by this density structure. The resulting temperature distribution, along with the assumption that the disk is in vertical hydrostatic equilibrium, are then used to calculate the vertical density of the disk and thereby obtain a new density structure. By repeatedly calculating the disk temperature and density, this algorithm converges to specify a self-consistent disk structure. For the sake of brevity, we will call this combined numerical model \RADp.

A given circumplanetary disk is characterized by a constant mass accretion rate $\dot{M}$, which is delivered onto a central planet of mass $M_p$. The disk viscosity follows the $\alpha$-prescription \citep{Shakura1973}, with $\alpha$ as a free parameter. The size of the disk is set by the magnetic truncation radius $R_X$ at the inner edge and the centrifugal radius $R_C$ at the outer edge. For a given planet radius $R_p$ and planetary surface magnetic field strength $B_p$, the truncation radius has the form \citep{Ghosh1978, Blandford1982}
\begin{equation}\label{eq:RXdef}
\begin{split}
    \frac{R_X}{R_p}\simeq&3.8\left(\frac{M_p}{\unit{M_J}}\right)^{-1/7}\left(\frac{\dot{M}}{\qty{10}{M_J\per\mega yr}}\right)^{-2/7}\\
    &\times\left(\frac{B_p}{\qty{500}{G}}\right)^{4/7}\left(\frac{R_p}{\qty{e10}{\centi\meter}}\right)^{5/7}\,.
\end{split}
\end{equation}
In this case, we will assume that
\begin{equation}
    B_p\simeq\qty{500}{G}\,\left(\frac{\dot{M}}{\qty{10}{M_J/Myr}}\right)^{1/3}\,.
\end{equation}
Although in practice the relationship between $R_X$ and $\dot{M}$ has significant scatter (e.g., \citealt{Thanathibodee2023, Pittman2025}), Eq. \eqref{eq:RXdef} provides a reasonable parameterization for these models. The centrifugal radius, which defines the outer boundary of the disk \citep{Quillen1998, Martin2011}, has the form 
\begin{equation}
    R_C=\frac{R_H}{3}=\frac{a}{3}\left(\frac{M_p}{3M_\star}\right)^{1/3}\,,
\end{equation}
where $R_H$ is the Hill radius, $a$ is the planet's orbital semimajor axis, and $M_\star$ is the mass of the host star.\footnote{This form for $R_C$ assumes that the angular momentum of the incoming material is determined by the orbital angular velocity at the Hill radius. This assumption is generalized below.}

The disk is assumed to reach a steady state and is heated by both viscosity and radiation from the central planet. The viscous heating is parameterized as an internal heat source $\Gamma$, which is given by \citep{Pringle1981, Frank2002}
\begin{equation}\label{eq:vischeat}
    \Gamma=\frac{9}{8}\alpha\rho v_s^2\Omega\,.
\end{equation}
In Eq. \eqref{eq:vischeat}, $\alpha$ is the viscosity parameter, $v_s=\sqrt{k_BT/\mu m_p}$ is the sound speed, and $\Omega=\sqrt{GM_p/R^3}$ is the (Keplerian) orbital frequency. The temperature $T$ is initially set to be the disk temperature calculated by \DIAD, but is later updated as the \RADp algorithm converges. 

The disk is also heated by the central planet, which is assumed to emit a luminosity equal to the energy delivered by the accreted material, so that the internal luminosity of the planet is negligible. The total luminosity of the planet is thus given by 
\begin{equation}\label{eq:Lp}
    L_p=\frac{GM_p\dot{M}}{R_p}\left(1-\frac{R_p^3}{3R_X^3}\right)\left(1-f_d\frac{R_p}{R_X}\right)\,,
\end{equation}
where $f_d$ is the fraction of infalling material accreted directly onto the disk. The second term in Eq. \eqref{eq:Lp} accounts for the planet's rotational energy and the third term accounts for the fact that the accreted material falls onto the planet from the disk's inner edge after some fraction of the luminosity has been lost to radiation and rotation in the disk. The corresponding effective surface temperature of the planet is 
\begin{equation}\label{eq:Tp}
    T_p=\left(\frac{L_p}{4\pi\sigma R_p^2}\right)^{1/4}\,.
\end{equation}

The emitted spectrum of the planet is modeled with the \texttt{SONORA Bobcat} model grid \citep{Marley2021}, which provides the spectra of planetary-mass objects for a grid of effective temperatures $T_{\rm eff}=\qtyrange{200}{2400}{K}$ and surface gravities $g=\qtyrange{10}{3160}{m\per s^2}$. Given a planet mass and radius, the surface gravity is $g=GM_p/R_p^2$, while Eq. \eqref{eq:Tp} provides the effective temperature. The \texttt{SONORA Bobcat} spectra are then interpolated between the temperature--surface gravity grid points to match the modeled planet, under the assumption that these planets have a solar metallicity. The \texttt{SONORA Bobcat} spectra are provided at $\sim\num{360000}$ wavelengths from $\lambda=\qtyrange{0.5}{50}{\micro\meter}$, which are downsampled to the \num{500} wavelength points used in our calculation. Outside of the wavelength range provided by the model grid, the planet's spectrum is set to be a blackbody with a temperature given by Eq. \eqref{eq:Tp}. This spectrum is then loaded into \RAD as the spectrum of the central planet. While this spectrum is fixed in these models, the planet will necessarily be heated by the radiation of the disk. Throughout this work, this ``backreaction'' heating accounts for a few percent of the total, but it can be increased if the magnetic truncation radius is smaller.

Setting the planet's effective temperature to be $T_{\rm eff}=T_p$ implicitly assumes that the infalling material is immediately well-mixed into the planet's atmosphere, so that the entire planet surface emits at a single temperature. However, some fraction of the accretion luminosity will not be distributed across the planet's surface but released in the accretion shock onto the planet. Appendix \ref{app:shocks} discusses the effects of these shocks on the radiative signatures. {This model also assumes that the stellar radiation is negligible, since the planet will be near the midplane of the circumstellar disk and thus be shielded from the star. Once the disk dissipates and stellar radiation becomes important, the mass accretion rate onto the protoplanet will be significantly reduced and the internal luminosity of the planet will become important. These models are not applicable in this scenario. }

\begin{figure}
    \centering
    \includegraphics{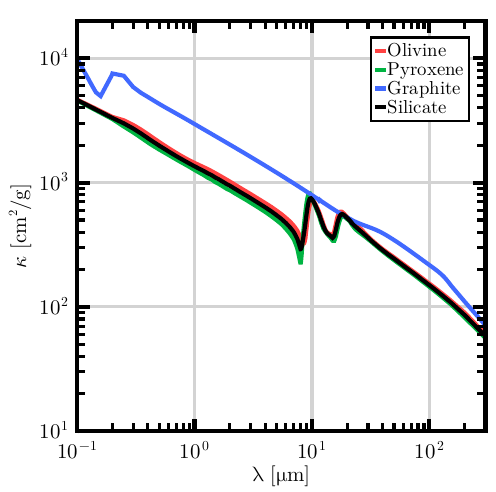}
    \caption{\textbf{Dust Opacities.} The dust opacities for different compositions. The silicate component of the dust (black) is composed of \qty{50}{\percent} olivine (red) and \qty{50}{\percent} pyroxene (green). The carbon component is entirely graphite (blue). The total opacity of the dust is the sum of the silicate and carbon components, with the silicate opacity weighted by the silicate fraction $f_{\rm Si}$ and the carbon opacity weighted by $(1-f_{\rm Si})$. The dust grains have a power-law size distribution $n(a)\propto a^{-3.5}$ with a minimum size of \qty{0.005}{\micro\meter} and a maximum size of \qty{100}{\micro\meter}. Note that the opacity is given in terms of per unit mass of dust. }
    \vspace{-10pt}
    \label{fig:opacomp}
\end{figure}

In the \RADp model, the dust is modeled as a single population throughout the disk, with a power-law size distribution so that $n(a)\propto a^{-3.5}$ between a minimum size $a_{\rm min}=\qty{0.005}{\micro\meter}$ and a maximum size $a_{\rm max}=\qty{100}{\micro\meter}$ {(following the classic distribution of \citealt{Mathis1977})}. The dust is also characterized by two global parameters --- the dust-to-gas mass ratio $\eta$ and the silicate fraction $f_{\rm Si}$. In this model, the dust is composed of a combination of olivine, pyroxene, and graphite, {with the opacity of each species given by \citet{DAlessio2001} and \citet{Micolta2024}}. The silicate component consists of \qty{50}{\percent} each olivine and pyroxene, while graphite is the sole constituent of the carbon component. If the effective (absorption plus scattering) opacities of these species are $\kappa_o$, $\kappa_p$, and $\kappa_g$ respectively, then the opacity per gram of dust is set to be
\begin{equation}
    \kappa=\frac{f_{\rm Si}}{2}\kappa_o+\frac{f_{\rm Si}}{2}\kappa_p+(1-f_{\rm Si})\kappa_g\,.
\end{equation}
These opacities are shown in Fig. \ref{fig:opacomp}. These opacities already account for the grain size distribution of the dust. Multiplying the opacity $\kappa$ by the dust-to-gas ratio $\eta$ gives the opacity per unit mass of the disk. {In this calculation, scattering is included by adding the the scattering opacity to the absorption opacity, so that the effective opacity is the sum of both. This approximation has negligible importance on either the disk structure or the radiative signatures shown in this paper. Scattering is only important when the system is viewed from near the equatorial plane, but the circumstellar disk will strongly attenuate the radiation in this viewing geometry and render detections impossible. }

Given the heat sources of viscosity and the planet's radiation and the well-defined opacity of the dust, we use \RAD to calculate the temperature structure of the disk. Assuming that the disk is in vertical hydrostatic equilibrium so that 
\begin{equation}
    \frac{\partial P}{\partial z}=-\frac{GM_p}{R^3}z\rho\,, \label{eq:HEQ}
\end{equation}
which is valid in the limit of small distances from the midplane, the disk temperature then determines the disk density structure. Assuming that the disk follows the ideal gas law with a fixed mean molecular weight $\mu$, so that $P=(\rho k_BT)/(\mu m_p)$ and defining $A=(GM_p\mu m_p)/(R^3k_B)$ for convenience, the condition of hydrostatic equilibrium becomes
\begin{equation}
    \frac{1}{\rho}\frac{\partial\rho}{\partial z}=-\left(\frac{Az}{T}+\frac{1}{T}\frac{\partial T}{\partial z}\right)\,.
\end{equation}
After integrating both sides, the density at a given radius $R$ can be written in the form  
\begin{equation}
\rho(z)=\rho(0)\exp\left[-A\!\!\int\displaylimits_0^z\!\!\frac{z\de z}{T(z)} - \ln\left(\frac{T(z)}{T(0)}\right)\right]\,. \label{eq:rhoint}
\end{equation}
Since the temperature $T(z)$ is known (from \RAD), this equation can be numerically integrated to find $\rho(z)$. 

The only remaining unknown is central density $\rho(0)$, which can be found by specifying the energy emitted per unit area (see \citealt{Armitage2020}{, accounting for the fact that the disk has two sides}) according to 
\begin{equation}
    D(R)=\frac{3}{8\pi}\dot{M}\Omega^2,
\end{equation}
ignoring the viscous boundary conditions. The energy released by viscous heating in this annulus takes the form 
\begin{equation}    
    D(R)=\int\displaylimits_0^\infty\!\!\Gamma(R,z)\,\de z =\frac{9}{8}\frac{\alpha\Omega k_B}{\mu m_p}\int\displaylimits_0^\infty\!\!\rho(z)T(z)\,\de z\,,
\end{equation}
where $\Gamma(z)$ is given by Eq. \eqref{eq:vischeat}. {The bounds of this integral are set so that the luminosity released in the upper half of the disk is equal to the luminosity emitted by the corresponding area.} Since the temperature is known, $\rho(z)=\rho_0f(z)$ and the above expression can be inverted to obtain 
\begin{equation}
    \rho_0=\frac{\dot{M}\Omega}{6\pi\alpha}\frac{\mu m_p}{k_B}\left[\int_0^\infty \!\!\!\!\!\!f(z)T(z)\de z\right]^{-1}\,,\label{eq:rho0int}
\end{equation}
where $f(z)$ is given by Eq. \eqref{eq:rhoint}. For each annulus in the disk, we numerically integrate Eq. \eqref{eq:rho0int} to calculate $\rho_0$ and the updated disk density distribution $\rho(R,z)$. 

\setlength{\tabcolsep}{2pt}
\begin{table}[t]
    \caption{\textbf{Canonical {Fiducial} Values.} {Fiducial} values for model parameters, which are repeatedly used throughout this paper. }
    \centering
    \begin{tabular}[t]{rcl}
        Variable & Symbol & Value \\\hline
        Stellar mass & $M_\star$ & \qty{1}{M_\odot} \\
        Orbital distance & $a$ & \qty{5}{au} \\
        Mass accretion rate & $\dot{M}$ & \qty{10}{M_J\per\mega yr} \\
        Viewing angle & $\psi$ & 0; polar\\
        Planet mass & $M_p$ & \qty{1}{M_J} \\
        Planet radius & $R_p$ & \qty{1.4}{R_J} \\
        Planet magnetic field & $B_p$ & \qty{500}{G} \\
        Infall geometry & - & isotropic \\
        Silicate fraction & $f_{\rm Si}$ & \num{1.0} \\
        Dust-to-gas ratio & $\eta$ & \num{0.0065} \\
        Viscosity parameter & $\alpha$ & 0.1 \\
    \end{tabular}
    \label{tab:canonvals}
\end{table}

This disk density is then loaded back into \RAD and the internal heating is updated to use the newly-calculated temperature distribution. By repeating the process of calculating the disk temperature and density, this procedure rapidly converges to a self-consistent disk structure. After only five iterations, the median fractional change in the density for a fiducial disk (see Table \ref{tab:canonvals}) is $\delta\rho/\rho\sim1\%$ {and is taken as converged}. The system SEDs are {then} calculated by the \RAD ray-tracing spectral module. {Additional iterations do not significantly change the radiative signatures or the structure of the disk, and Appendix \ref{app:conv} demonstrates that this model accurately reproduces the temperature distribution and radiative signatures of a thin, viscously-heated disk where appropriate.} 

In addition, the radiation from the disk and the planet are not viewed directly but are modified by the circumplanetary envelope, which is determined by the accretion of material through the Hill sphere. We follow the treatment of \citet{Taylor2024}, which we briefly summarize here. The infalling material is characterized by a polar asymmetry function $f_i$ defined such that the net mass inflow rate at a polar angle $\theta_0$ on the outer boundary is $\dot{M}f_i(\theta_0)$. These functions were set to have the form 
\begin{equation}
    f_i(\theta_0)=\begin{cases}
    3\cos^2\theta_0 & \text{polar;}\\
    1 & \text{isotropic;}\\
    \tfrac{3}{2}(1-\cos^2\theta_0) & \text{equatorial.}
    \end{cases}
\end{equation}
For ballistic infall, the material moves on zero-energy Keplerian orbits, which specifies the velocity of the incoming material at each point. The conservation of mass along streamlines then specifies the envelope density field throughout the circumplanetary envelope. The functional form of the density has been given by many previous authors (see, e.g., \citealt{Ulrich1976, Cassen1981, Chevalier1983, Adams2022, Adams2025}, and many others) and will not be repeated here. Note that the assumption of ballistic infall is valid if and only if the envelope is able to efficiently cool. However, if the envelope is unable to cool and hydrodynamical pressure effects become important, then the system is generally unable to collapse to form either a planet or a circumplanetary disk (e.g., \citealt{Krapp2024}). {Previous work using \RAD to model the temperature structure of these envelopes \citep{Taylor2025} has shown that the envelope is optically thin to its own radiation for $\dot{M}\lesssim\qty{30}{M_J/Myr}$, implying that the envelope can cool sufficiently for these mass accretion rates.}

For a given (converged) disk model, the envelope is added by setting the density at every point within the Hill radius to be the maximum of the envelope density (scaled by the dust-to-gas ratio) and the \RADp-calculated disk density. The \RAD code is then used to calculate the temperature of the dust in both the disk and the envelope, and thus accounts for both viscous heating in the disk and the radiation of the central planet. Note that up until this point, the models have been agnostic to the infall geometry, since the steady-state disks are nearly independent of this parameter \citep{Taylor2024, Adams2025}. However, the infall distribution will heavily influence the envelope density structure. 

\begin{figure*}[t!]
    \centering
    \includegraphics{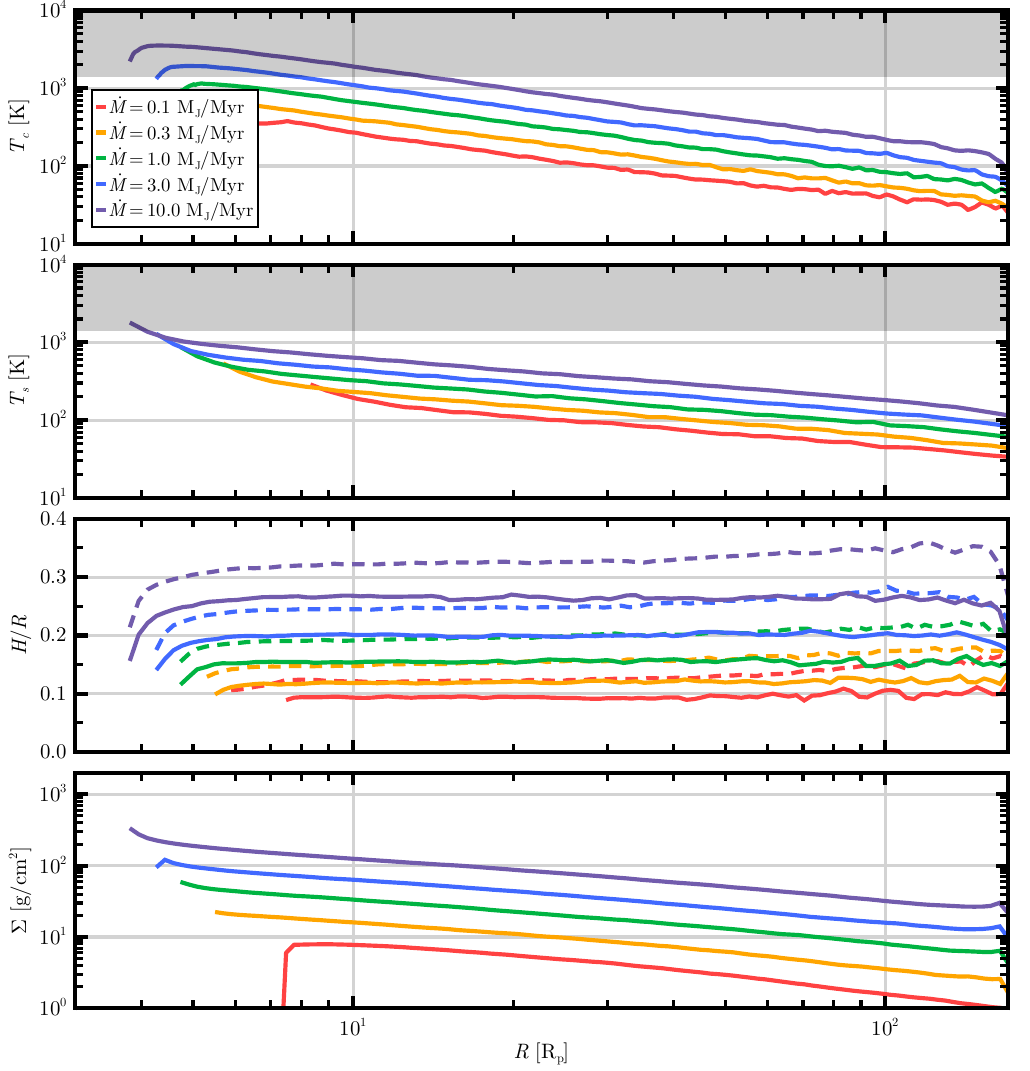}
    \caption{\textbf{Disk Radial Structure.} The central temperature, surface temperature, aspect ratio, and surface density versus radius for different mass accretion rates. The mass accretion rates are shown in different colors. In the top two panels, the gray region shows where the dust sublimates at $T\geq\qty{1400}{K}$. The variations in the disk aspect ratio at large distances results from Monte Carlo noise in the temperature. The dashed lines in the third panel show the scale height if the disk were isothermal at the central temperature. The {total} masses of these disks are $M_p=\qtylist{7.9e-6; 1.8e-5; 4.0e-5; 8.1e-5;1.7e-4}{M_J}$ for $\dot{M}=\qtylist{0.1;0.3;1;3;10}{M_J\per\mega yr}$, respectively. }
    \vspace{-10pt}
    \label{fig:rad_struc}
\end{figure*}

\section{Disk Structure}\label{sec:diskstruc}

This section discusses the disk structure calculated by the \RADp model and shows how this structure varies for mass accretion rates ranging from $\dot{M}=\qtyrange{0.1}{10}{M_J/Myr}$. Here, and unless otherwise noted, the system parameters are set to fiducial values, which are given in Table \ref{tab:canonvals}.

The fiducial values of these parameters are selected to approximate the expected values of a young Jupiter. However, the specification of the mass accretion rate $\dot{M}$ is less certain. Since disk lifetimes are measured in Myr and giant planet masses are comparable to Jupiter, the mass accretion rate $\dot{M}\sim\qty{1}{M_J\per\mega yr}$ represents a typical value (e.g., \citealt{Helled2014,Adams2021}). However, planets have been observed that are as large as \qty{10}{M_J}. To produce such a planet within \qty{1}{\mega yr}, $\dot{M}$ must have an average value closer to \qty{10}{M_J\per\mega yr}. Note that forming planets are more easily detected with larger mass accretion rates, as the system luminosity scales with $\dot{M}$. In any case, we set the fiducial value to $\dot{M}=\qty{10}{M_J\per\mega yr}$ and show how the disk structure and radiative signatures vary with the mass accretion rate. 

\subsection{Radial Structure}

We first consider the radial structure of the disk. Fig. \ref{fig:rad_struc} shows the central temperature {$T_c$}, the disk surface temperature {$T_s$}, the aspect ratio $H/R$, and the surface density of the disks versus radius. Some definitions are necessary here. The disk surface temperature $T_s(R)$ is defined to be $T(R,z_s)$, where $z_s$ is the vertical height such that the radial optical depth into the disk $\tau_p$ is unity. At a polar angle $\theta$ and radius $r$ in the disk,\footnote{The lowercase $r$ is defined to be the radial distance to the origin, while the uppercase $R$ is the distance in the $x$-$y$ plane.} this optical depth is defined to be
\begin{equation}\label{eq:taurad}
    \tau_p=\int_0^r\!\!\!\kappa_R(\rho,T_p)\rho(r,\theta)\,\de r\,.
\end{equation}
At each annulus, $z_s$ is found by iteratively integrating deeper into the disk until $\tau_p\geq1$. This optical depth characterizes the opacity of the disk to the radiation from the central planet.

In addition, the scale height $H$ is not precisely defined, since the disks are not vertically isothermal and the density distribution is not an exponential function of the height. For our purposes, the scale height will be calculated as the height $H$ at which $\rho(H)=\rho_0\exp(-0.5)$, where $\rho_0$ is the central density. This condition is true for vertically isothermal or exponentially decaying disks and so is an appropriate definition in these circumstances. For the sake of comparison, Fig. \ref{fig:rad_struc} also shows the scale height if the disk was isothermal with a temperature equal to the central temperature, so that $H=\sqrt{2k_BT_c/(\Omega^2\mu m_p)}$.

Inspecting Fig. \ref{fig:rad_struc} reveals that disks with larger mass accretion rates are hotter, since the total disk luminosity scales with $\dot{M}$ and higher luminosities require higher temperatures to achieve energy balance. In fact, for mass accretion rates $\dot{M}\geq\qty{1}{M_J\per\mega yr}$, \RADp predicts that the central temperature in the inner disk will be significantly higher than the sublimation temperature. This behavior is a consequence of the fact that \RADp does not account for the changes in opacity due to dust sublimation. In reality, the opacity will be approximately zero in this region and therefore no temperature gradient is necessary to transport the energy, setting an effective maximum temperature at \qty{1400}{K}. However, this unphysically hot region is deeply embedded in the disk and will have negligible effect on the radiative signatures of the system. 

In models with a sufficiently high mass accretion rate, the inner edge of the disk is cooler than nearby annuli farther from the central planet. This reversed temperature scaling is a consequence of radial {energy} transport. At some point in the inner regions of the disk, the radial optical depth to the inner wall becomes smaller than the vertical optical depth. Since the central temperature is primarily set by the minimum optical depth to infinity, the disk at the inner edge cools more efficiently and has a lower temperature than annuli at slightly larger radii. While the planet's radiation will also be most significant in this region, the high optical depth of the disk means that the equilibrium temperature can slightly increase with radius. 

The second panel shows that the disk surface temperatures {(the temperature where $\tau_p=1$)} are colder than the central temperatures, as expected. Since the disks are optically thick, the surface temperature must be lower than the central temperature, imposing a temperature gradient that radiatively transports the viscous luminosity to the disk surface. While the surface temperature generally follows a power-law scaling with the radius (especially at large disk radii), the surface temperature increases significantly at the inner edge of the disk. This increase is a consequence of the heating from the central planet, which becomes significant at the inner edge of the disk. 

Now we consider the aspect ratio of the disk. First, disks with higher mass accretion rates are geometrically thicker, a consequence of the increased optical depth and viscous luminosity in these systems. Once again, the disk's scale height drops at the inner edge as the radial optical depth lowers and the cooling becomes more efficient, lowering the temperature and the disk scale height. Finally, the aspect ratios become roughly constant at large radii, independent of mass accretion rate, implying that the average disk temperature $T\propto R^{-1}$. Since the central temperature is the maximum temperature in a given annulus and $H\propto T^{1/2}$, the scale heights calculated using the central temperature are strictly larger than the scale height defined in terms of the density.

As discussed in previous work, the surface densities have a profile similar to $\Sigma\propto r^{-1}$, although they drop off steeply at the inner and outer boundaries. In addition, while higher mass accretion rates imply higher surface densities, the surface density does not scale linearly with the mass accretion rate. At higher mass accretion rates, the disk is necessarily hotter, meaning that for a fixed value of $\alpha$ the total viscosity of the disk will increase. A larger viscosity will more effectively transport material onto the planet, so the surface density of the disk will be a sublinear function of the mass accretion rate --- i.e., if $\Sigma\propto\dot{M}^n$, the index $0<n<1$.  

The temperature dependence of the viscosity has particularly significant implications when comparing the \RADp model to a geometrically razor-thin disk---that is, a disk that is confined to the equatorial plane with zero vertical extent. In a given annulus, both disks will have similar luminosities and therefore similar effective temperatures. However, the razor-thin disk will have a single temperature in the annulus by construction, while the geometrically thick disk will have an internal temperature that is higher than the emission temperature by a power of the disk optical depth. For identical system parameters, the thick disk will thus have a larger effective viscosity and therefore a smaller surface density in a given annulus. This effect means that complex two-dimensional models are necessary to accurately predict the masses and optical depths of these disks, both of which will be smaller than expected from razor-thin analytical models. 

\begin{figure}
    \centering
    \includegraphics{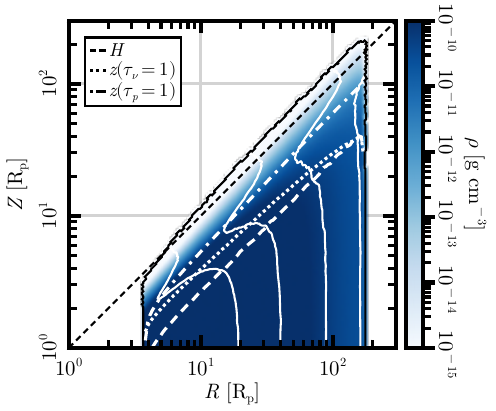}
    \caption{\textbf{Disk Density Structure.} The disk density (dust and gas) as calculated by our \RADp model. The black line indicates the ``boundary'' of the disk, outside of which the density is zero by construction. The white dashed line shows where $Z=H$, the white dotted line shows where the vertical optical depth to viscous heating $\tau_\nu=1$, and the white dash-dot line shows where the radial optical depth to the planet $\tau_p=1$. The system parameters are given by the fiducial values (Table \ref{tab:canonvals}). The solid white lines are isotherms at $T=\qtylist{100;250;500;1000}{K}$. The black dashed line shows where $Z=R$.}
    \vspace{-10pt}
    \label{fig:rhofig}
\end{figure}

\subsection{Vertical Structure}

We now consider the vertical structure of the disk. Fig. \ref{fig:rhofig} shows the density structure of the fiducial \RADp model. This figure demonstrates that {since} the disk aspect ratio is $H/R\sim0.25$, the disk is geometrically thick, with somewhat large densities relatively close to the line $Z=R$. In addition, the disk closely matches the density structure predicted by viscous heating alone. 

Fig. \ref{fig:rhofig} also shows the scale height $H$ and the surfaces where the optical depths are unity. Since there are two sources of heating in the disk (viscosity and {irradiation from} the central planet), there are two such surfaces, corresponding to where the internal radiation field ({mostly} produced by viscous heating) will easily escape and where the planet's radiation can easily penetrate. There are two corresponding optical depths, which will be equal to unity at these surfaces. Since the internal radiation field is generally transported vertically within the disk, the first optical depth ($\tau_\nu$) is defined to be the optical depth between $z=\infty$ and $z$, where the Rosseland mean opacity is weighted by the disk's local temperature. This optical depth is given by 
\begin{equation}\label{eq:tauvis}
    \tau_\nu\equiv\int_{z_s}^\infty\!\!\!\!\kappa_R(\rho,T_d(R,z))\rho(R,z)\,\de z\,.
\end{equation} 
This definition reflects the fact that the internal radiation at a point in the disk is approximately given by the blackbody function at the local temperature.

On the other hand, the planet's radiation is emitted from the system center at a fixed temperature $T_p$, so the planet optical depth $\tau_p$ is defined in terms of the radial coordinate. This optical depth is given by Eq. \eqref{eq:taurad}.
In contrast to the disk optical depth, in Eq. \eqref{eq:taurad} the Rosseland mean opacity is weighted by the blackbody corresponding to the emission temperature of the planet $T_p$, rather than the local disk temperature. For a given polar angle $\theta$, the radius $r_s$ where $\tau_p=1$ is found by numerically solving Eq. \eqref{eq:taurad} for where $\tau_p=1$. The set of $(\theta,r)$ points that correspond to $\tau_p=1$ are also shown in Fig. \ref{fig:rhofig}. The region between these two $\tau=1$ surfaces will necessarily be at a lower temperature than the immediately surrounding regions. In this region, $\tau_p\gg1$ and $\tau_\nu\ll1$, so the planet's radiation will be unable to penetrate but the radiation from viscous heating will easily escape into space. As a result, this region will experience weaker heating than the rest of the disk and will have a temperature minimum, similar to other irradiated accretion disks, such as T Tauri stars \citep{DAlessio1998, DAlessio1999, DAlessio2001, DAlessio2006}.  

\begin{figure*}
    \centering
    \includegraphics{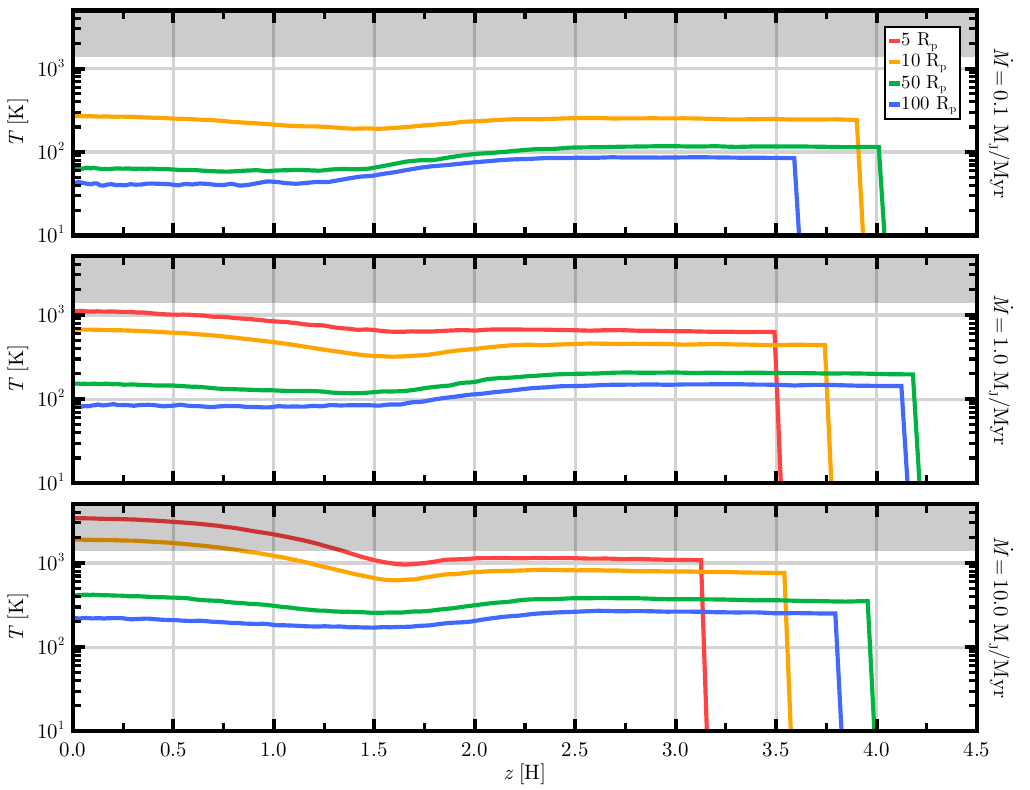}
    \caption{\textbf{Disk Temperature Structure.} The vertical temperature structure of the disk at \qtylist{5;10;50;100}{R_p}. The height is shown in units of the (isothermal) scale height, defined to be $H\equiv\sqrt{(2k_BT_c)/(\Omega^2\mu m_p)}$, where $T_c$ is the central temperature and $\Omega=\sqrt{GM_p/R^3}$ is the Keplerian angular orbital frequency. For $\dot{M}=\qty{0.1}{M_J\per\mega yr}$, $R_X>\qty{5}{R_p}$, and so that line does not appear. The gray region indicates where $T\geq\qty{1400}{K}$ and where dust will sublimate. }
    \vspace{-10pt}
    \label{fig:temp_comp}
\end{figure*}

\begin{figure}
    \centering
    \includegraphics{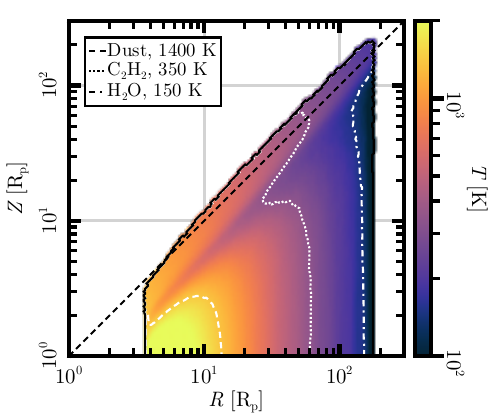}
    \caption{\textbf{Disk Temperature Structure.} The disk density (dust and gas) as calculated by our numerical \RADp model. The black line indicates the ``boundary'' of the disk, outside of which the density is zero. The system parameters are given by the fiducial values (Table \ref{tab:canonvals}). The white lines show the snowlines of dust, C$_2$H$_2$, and H$_2$O. }
    \vspace{-10pt}
    \label{fig:disk_temp}
\end{figure}

Fig. \ref{fig:temp_comp} shows the vertical temperature structure of the disk at several radii and a range of mass accretion rates. Several notable effects are visible in these plots. First, the central temperature in the inner regions of these disks can reach $T\geq\qty{1400}{K}$, above the dust sublimation temperature. This result is a reflection of the fact that the model does not account for the dust sublimation. Second, at the inner edge of the disk and for high mass accretion rates, the temperature has a small minimum at around $z=1.75H$, with the temperature flattening out above this point. As previously discussed, this region arises because of the mismatch between two different $\tau=1$ surfaces --- the $\tau_p$ to the planet in the radial direction and $\tau_\nu$ in the vertical direction. 

Interior to this point, $\tau_p\gg1$ and the disk {temperature is dominated by the viscous heating}. At the same time, the viscous heating is primarily produced in this region and must be transferred outwards. Since $\tau_\nu\gg1$, this transport requires a steep temperature gradient until the $\tau_\nu=1$ surface, where the radiation freely escapes the disk and the temperature no longer increases. However, once $\tau_p\ll1$, the planet's radiation will heat the disk and raise the temperature slightly. In the region where $\tau_\nu\ll1\ll\tau_p$, heating from the planet is minimal while radiation from viscous heating can efficiently escape, producing a temperature minimum. 

At larger disk radii and smaller mass accretion rates, the temperature has a vertical inversion and is constant towards the disk midplane (Fig. \ref{fig:temp_comp}). In this region, $\tau_\nu\leq1$ even at the midplane, so that the entire disk is optically thin to the viscous heating and no temperature gradient is necessary to transport the energy. At the same time, the geometry of the disk results in a well-defined $\tau_p=1$ surface, above which the planet's radiation heats the disk above the central temperature.  It is also worth noting that the $\tau_p\ll1$ region heated by the planet alone is far larger in volume than the $\tau_\nu\gg1$ region where viscous heating dominates, which is a consequence of both the geometry of the disk and the fact that the upper regions of the disk are at a relatively low density. However, most of the disk mass will be near the disk midplane where $\tau_\nu\gg1$. 

Fig. \ref{fig:disk_temp} shows the \RADp temperature structure of a fiducial disk. The temperature minimum region shown in Fig. \ref{fig:temp_comp} is clearly visible in this plot and bounded by the $\tau=1$ surfaces (see Fig. \ref{fig:rhofig}). This figure also shows where the temperature of the disk is equal to the approximate sublimation temperature of several different critical molecules --- specifically, where dust sublimates (\qty{1400}{K}), refractory organics become C$_2$H$_2$ (\qty{350}{K}, \citealt{Houge2025}) and H$_2$O sublimates (\qty{150}{K}, \citealt{Tobin2023}). The minimum temperature of the disk is above the sublimation temperatures of both CO (\qty{20}{K}, \citealt{Harsono2015}) and CO$_2$ (\qty{50}{K}, \citealt{Harsono2015}), so these molecules will be gaseous throughout the disk.

The innermost regions of this fiducial disk will have sublimated silicates, so the dust disk will be effectively cut off at a slightly larger radius than the magnetic truncation radius. In addition, C$_2$H$_2$ will only be gaseous in the inner regions of this fiducial disk, and complex organic ``soot'' will be present throughout the remainder of the disk, which accounts for approximately \qty{83}{\percent} of the disk mass. H$_2$O ice will be limited to the outer regions of the disk, although this icy region accounts for \qty{28}{\percent} of the total mass of the disk. The phases of these molecules, as well as CO and CO$_2$, may have significant impacts on the spectra, since the gaseous volatiles will add spectral lines while the solids may contribute to the dust opacity beyond these snowlines. The sublimation regions of these molecules will also vary significantly with the system parameters, which have a significant impact on the disk temperature structure. Incorporation of the effects of volatiles and their phase changes are left for future work. 

\subsection{Envelope Heating}\label{sec:envheat}

In our construction of the circumplanetary environment, we have assumed that the disk's density structure is completely independent of the {presence of the} envelope.\footnote{{Recall that the envelope density is set by assuming ballistic infall and is discussed in Sec. 2. (see \citealt{Ulrich1976} and citations in that section).}} In fact, the envelope will irradiate the disk's surface and thereby increase the scale height of the disk. To determine the importance of this effect, {we compare the temperature of the disk when the envelope is or is not included.} The difference between these temperatures is then divided by the maximum disk temperature when the envelope is not included to obtain $(\Delta T/T)_{\rm max}$. This envelope backreaction heating is most significant at high mass accretion rates, when the envelope is optically thick near the disk and the surface and the outer edges of the disk are unable to cool as efficiently, thereby increasing their temperature. In any case, the maximum change is $\sim\qty{10}{\percent}$ of the maximum disk temperature. 

This temperature change is accounted for when calculating the temperature of the system and the radiative signatures. The only effect of the envelope that is not included in the model is the increase in the disk scale height due to the hotter upper layers, which can change the density of the system. However, the density at each point is set to the maximum of the envelope and disk densities. If the disk scale height slightly increases and the density of the disk lowers, the envelope will replace the disk in this region and leave the actual density essentially unchanged. As a result, the (unaccounted-for) effects of envelope heating are generally minor. 

\begin{figure}
    \centering
    \includegraphics{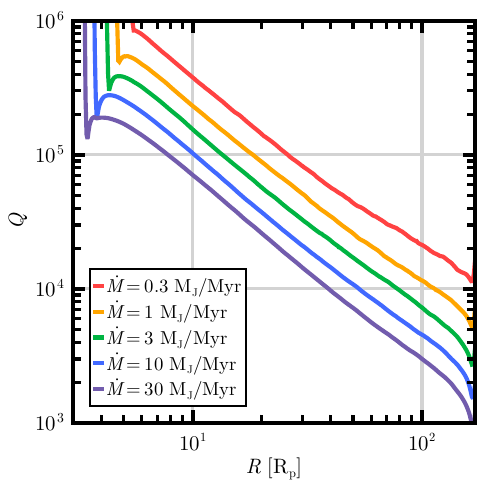}
    \caption{\textbf{Toomre $Q$ Parameter.} The Toomre $Q$ parameter versus radius and mass accretion rate. Except for the mass accretion rate, all parameters are set to their fiducial values. }
    \label{fig:Qvalue}
\end{figure}

\subsection{Gravitational Instability}\label{subsec:GI}

Another critical question is whether or not these circumplanetary disks are gravitationally unstable. If gravitational instability can operate, then it could transport material onto the central protoplanet. On the other hand, a gravitationally stable disk requires viscosity to transport material to the planet. 

A disk region is subject to gravitational instability when the Toomre parameter
\begin{equation}
    Q=\frac{v_s\kappa}{\pi G\Sigma}\lesssim1\,,
\end{equation}
where $v_s$ is the sound speed, $\kappa\simeq\Omega$ is the epicyclic frequency, and $\Sigma$ is the disk surface density \citep{Toomre1964}. Note that this condition applies to axisymmetric perturbations. Since $v_s=\sqrt{k_BT/\mu m_p}$ depends on the temperature and the disk is not vertically isothermal, the sound speed at a given annulus is ill-defined. We use a mass-averaged temperature $\overline{T}$ given by
\begin{equation}
    \overline{T}(R)\equiv\frac{2}{\Sigma}\int\displaylimits_0^\infty\!\! T(R,z)\rho(R,z)\,\de z
\end{equation}
so that $v_s(R)=\sqrt{k_B\overline{T}(R)/\mu m_p}$.

Fig. \ref{fig:Qvalue} plots the Toomre $Q$ parameter versus radius for fiducial disks over a range of mass accretion rates. These disks are gravitationally stable, with $Q\gtrsim\num{e3}$, so most of the mass accreted through the disk must be driven by viscosity and not by gravitational instability.  

\subsection{Thermal Instability}\label{subsec:TI}

While these disks are gravitationally stable, at a high mass accretion rate ($\dot{M}\gtrsim\qty{60}{M_J\per\mega yr}$) they are subject to the disk thermal instability, which has been discussed extensively in previous work (see, e.g., \citealt{Kawazoe1993, Bell1994, Frank2002} Ch. 5.8; \citealt{Hartmann2009}). At these mass accretion rates, a viscous disk has no steady-state solution. 

This instability occurs when the disk's central temperature becomes comparable to the ionization temperature of hydrogen. At this point, the opacity decreases with increasing temperature, reducing the disk's capacity to cool and causing a temperature runaway until the opacity once again has a positive temperature dependence and the temperature stabilizes. However, this stable temperature corresponds to a higher mass accretion rate through the disk than the incoming accretion rate onto the disk. The disk mass will become depleted until the equilibrium temperature state is below that of the unstable region. This new stable temperature will correspond to a net mass accretion onto the disk, with the disk mass increasing until the disk temperature once again jumps to the high value.\footnote{In circumstellar disks, this mechanism is suspected to be the cause of FU Orionis accretion outbursts \citep{Hartmann1996, Hartmann2009}.} These feedback processes indicate that a disk with temperatures in this regime does not have a steady-state solution.

This thermal instability generally sets in when a reasonable portion of the disk has an effective temperature $T_{\rm eff}\gtrsim\qty{2000}{K}$ \citep{Bell1994}. Ignoring the planet's radiation (which will generally deliver less energy to the disk than the viscous heating), the effective temperature at a radius $R$ is given by \citep{Armitage2020}
\begin{equation}
    \sigma T_{\rm eff}^4=\frac{3GM_p\dot{M}}{8\pi R^3}\left(1-\sqrt{\frac{R_p}{R}}\right)\,.
\end{equation}
If $R =R_X\simeq 3R_p$, the condition for the thermal instability to set in ($T_{\rm eff}\geq\qty{2000}{K}$) becomes 
\begin{equation}\label{eq:Mdotunstable}
    \dot{M}\geq\qty{63.7}{M_J\per\mega yr}\left(\frac{M_p}{\qty{1}{M_J}}\right)^{-1}\,.
\end{equation}

The calculation of the disk's thermal and radiative structure relies on the assumption of a steady-state disk as an initial condition. The above results imply that the disk structure cannot be calculated if $\dot{M}\gtrsim\qty{60}{M_J\per\mega yr}$. Since \RADp relies on \DIAD to calculate the initial structure of these disks, and \DIAD accounts for the ionization processes necessary for the thermal instability, neither model converges for mass accretion rates beyond this critical value. The thermal instability thus sets a maximum for both the mass accretion rate we can effectively calculate and for the average mass accretion rate these systems could attain. {Note that for smaller and more realistic $\alpha$ values, the disk will have a higher surface density and central temperature. Therefore, the thermal instability may set in at lower mass accretion rates than calculated here.} Throughout this work, the mass accretion rate is held below the expected accretion rate ($\dot{M}=\qty{60}{M_J\per\mega yr}$) where thermal instability is expected. 

\section{Disk Radiative Signatures}\label{sec:disksig}

This section explores how the system parameter values affect the emergent radiative signatures, both including and excluding the effects of the circumplanetary envelope. 

\begin{figure}
    \centering
    \includegraphics{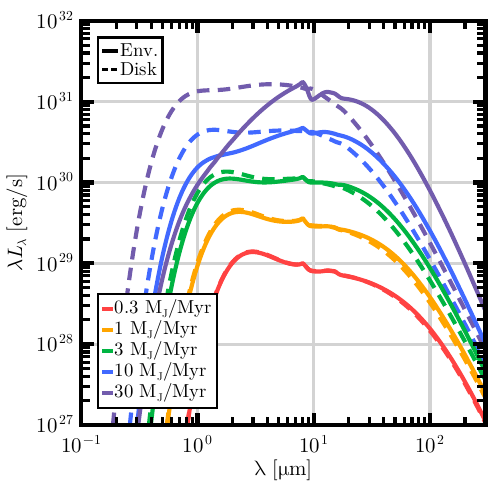}
    \caption{\textbf{Effects of Mass Accretion Rate.} The spectral energy distributions of the planet/disk/envelope system for a range of mass inflow rates. The system is assumed to be viewed from above. Aside from the mass inflow rate, the system parameters are set to their fiducial values (Table \ref{tab:canonvals}). The line style indicates whether the envelope has been included (solid) or not (dashed).}
    \vspace{-10pt}
    \label{fig:envgridMpd}
\end{figure}

\subsection{Mass Accretion Rate}\label{subsec:Mdotvar}

In Fig.~\ref{fig:envgridMpd}, we show the SEDs produced by systems at a range of mass accretion rates. As expected, the mass accretion rate strongly affects the radiative signatures, as higher mass accretion rates necessarily imply a higher system luminosity.

The high luminosity associated with a high mass accretion rate will necessarily lead to a geometrically thick disk. Since a thick disk will absorb more of the planet's radiation than a thin disk and reradiate it in the polar direction, the disk's luminosity will be shifted to preferentially be along the system pole relative to that of the planet. At this viewing angle, a thick disk will thus display a higher apparent disk-planet luminosity ratio, even though the ratio between the total luminosity emitted by the disk and planet is approximately constant. 

At low mass accretion rates, the total spectrum is relatively smooth, with a small \qty{10}{\micro\meter} silicate absorption feature. As the mass accretion rate and envelope optical depth increase, the \qty{10}{\micro\meter} silicate feature becomes more prominent. By $\dot{M}=\qty{30}{M_J\per\mega yr}$, this feature is a significant component of the spectrum around $\lambda=\qty{10}{\micro\meter}$. This increased prominence is a natural consequence of the increased envelope optical depth, since more dust will necessarily absorb more radiation and produce a stronger spectral feature.

At the same time, at high mass accretion rates the spectrum shifts closer to that of a blackbody. This shift is a consequence of the increased optical depth of the envelope, which absorbs short-wavelength radiation from the planet and reprocesses it to longer wavelengths. Although there will always be a range of temperatures within the system, if the optical depth of the envelope becomes much greater than unity, the envelope develops an effective photosphere that emits much like a blackbody. As a result, increasing the optical depth pushes the system SEDs towards this limit. The increased envelope optical depth at higher mass accretion rates can be observed by comparing the solid and dashed lines in Fig. \ref{fig:envgridMpd}, which show the SED with and without the envelope. While the envelope has essentially no effect for $\dot{M}\leq\qty{1}{M_J\per\mega yr}$, it significantly modifies the spectrum at $\dot{M}=\qty{30}{M_J\per\mega yr}$ as the envelope density scales with the mass accretion rate.

\begin{figure}
    \centering
    \includegraphics{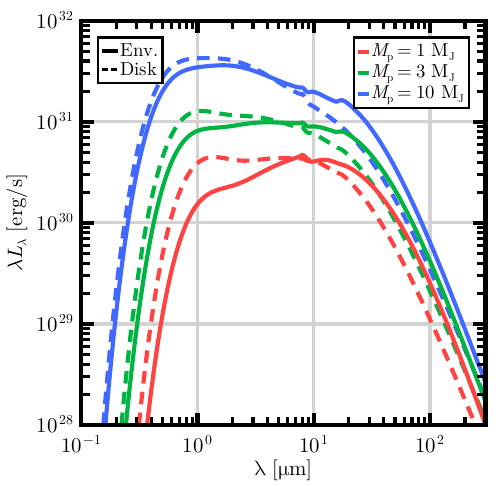}
    \caption{\textbf{Effects of Planet Mass.} The spectral energy distributions of the planet/disk/envelope system for a range of planet masses. The system is assumed to be viewed from above. Aside from the planet mass, the system parameters are set to their fiducial values (Table \ref{tab:canonvals}). The line style indicates whether the envelope has been included (solid) or not (dashed).}
    \vspace{-10pt}
    \label{fig:envgridMp}
\end{figure}

\subsection{Planet Mass}

Fig. \ref{fig:envgridMp} shows the effects of the planet mass on the spectral energy distributions. High-mass planets have deeper potential wells and therefore have larger system luminosities. However, the envelope column density (and thus optical depth) scales as $N_{\rm col}\propto M_p^{-2/3}$ \citep{Adams2022}. This scaling means that increasing the planet mass reduces the effects of the envelope on the radiative signatures and suppresses the \qty{10}{\micro\meter} feature. 

In addition, the spectra from systems with higher-mass planets are shifted to short wavelengths relative to the fiducial spectrum. This effect is a consequence of disk geometry --- the increased gravity of high-mass planets makes their circumplanetary disks geometrically thinner than the fiducial disk. These thinner disks absorb less of the planet's radiation and emit across a larger range of viewing angles than a thicker disk, thereby reducing the relative luminosity of the disk when viewed from the system pole. The planet's short-wavelength emission thus increasingly dominates the spectrum as the planet mass increases.  

\begin{figure}
    \centering
    \includegraphics{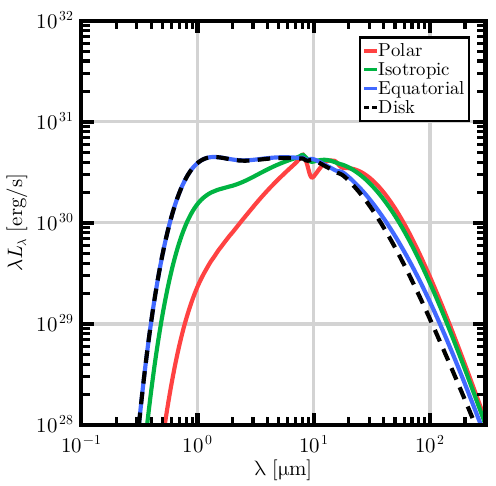}
    \caption{\textbf{Effects of Infall Function.} The effects of the infall on the spectral energy distributions of the planet/disk/envelope system. The system is assumed to be viewed from above. Aside from the accretion flow geometry, the system parameters are set to their fiducial values (Table \ref{tab:canonvals}). The disk-only spectrum is shown as the black dashed line and is not modified by the infall function.}
    \vspace{-10pt}
    \label{fig:envgridinfall}
\end{figure}

\subsection{Infall Geometry}

Since the disk is dominated by viscous accretion, the geometry of the infalling material has little effect on its surface density distribution and (unattenuated) radiative signatures. However, the infall function does significantly affect the envelope density structure and changes the envelope optical depth across viewing angles. The system SEDs for different infall functions are plotted in Fig. \ref{fig:envgridinfall}. We only show one disk-only spectrum, since the infall function has no effect on this SED.

The distinctions between the infall functions can be explained by the differences in the envelope optical depth to the observer. Since these SEDs are calculated for an observer along the system pole, the envelope optical depth increases for more polar infall geometries --- i.e., the optical depth is larger for polar infall than isotropic, and larger for isotropic infall than equatorial. At higher optical depths, the radiation from the planet and inner disk ($\lambda\lesssim\qty{8}{\micro\meter})$) will be increasingly absorbed and reprocessed into long-wavelength radiation from the envelope.

When the envelope infall is concentrated in the equatorial plane, it has little effect on the system SEDs. The planet is barely attenuated, and the only detectable effect is a minor increase in the long-wavelength SED as the envelope reprocesses small amounts of radiation. However, the envelope is increasingly heated by the planet and hot inner disk as the infall, and thus the envelope density, are concentrated towards the system pole. With a larger envelope column density along the line of sight, the envelope will significantly attenuate the short-wavelength radiation from the planet and disk and reprocess this radiation to longer wavelengths. The SED for an isotropic infall geometry clearly demonstrates this effect, which is only more significant for polar accretion. For a polar infall geometry, the envelope's optical depth is sufficiently high to produce a prominent dust absorption feature at \qty{10}{\micro\meter}. 

Notably, the significant variation in the radiative signatures with the infall geometry means that if an envelope is present, future observations will be able to constrain this parameter, albeit jointly with other system values. 

\begin{figure}
    \centering
    \includegraphics{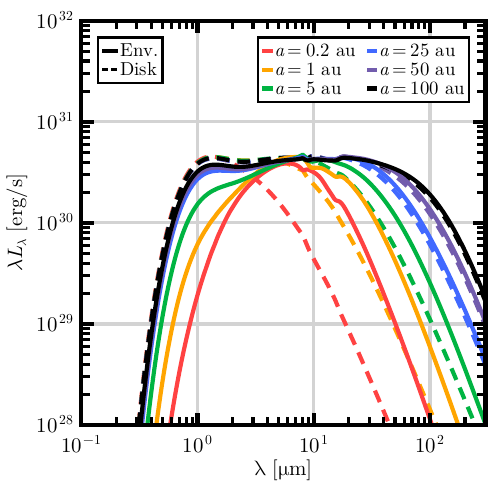}
    \caption{\textbf{Effects of Semimajor Axis.} The spectral energy distributions of the planet/disk/envelope system for a range of orbital semimajor axes. The system is assumed to be viewed from above. Aside from the semimajor axis, the system parameters are set to their fiducial values (Table \ref{tab:canonvals}). The line style indicates whether the envelope has been included (solid) or not (dashed).}
    \vspace{-10pt}
    \label{fig:envgrida}
\end{figure}

\begin{figure}
    \centering
    \includegraphics{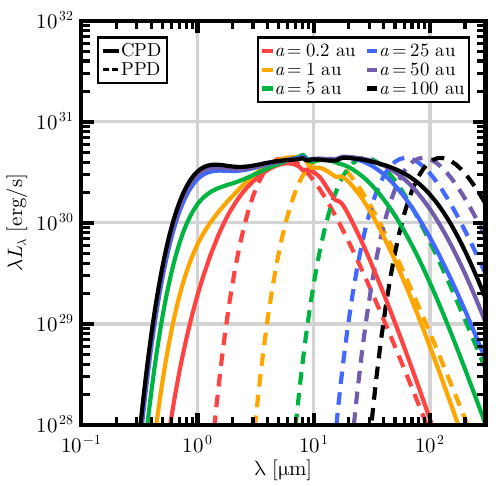}
    \caption{\textbf{{Background Disk Comparison.}} {The spectral energy distributions of the planet/disk/envelope system for a range of orbital semimajor axes. The system is assumed to be viewed from above. Aside from the semimajor axis, the system parameters are set to their fiducial values (Table \ref{tab:canonvals}). The SEDs of the portion of the background circumstellar disk that the planetary system replaces are shown as dashed lines. The solid lines show the SEDs of the system with the circumplanetary envelope included.}}
    \vspace{-10pt}
    \label{fig:PPDcomp}
\end{figure}

\subsection{Semimajor Axis}

Fig. \ref{fig:envgrida} shows how the system SEDs vary with the planet's orbital semimajor axis, which sets the size of the Hill radius/system boundary. {Note that the $a=\qty{100}{au}$ SED should be somewhat comparable to the \qty{300}{M_\oplus} curve from \citet{Choksi2025}, although our mass accretion rate is an order of magnitude larger and the system structure is somewhat different. Despite these differences, the breadth of the SED is approximately equivalent.} While the luminosity and spectrum of the planet in this model are unaffected by the Hill radius, the disks around planets orbiting closer to their star will be more significantly truncated by the stellar tide. Since the outer disk is cool, removing the outer disk lowers the long-wavelength parts of the disk-only spectrum. 

For a fixed mass accretion rate, a smaller Hill sphere results in a denser envelope, which increasingly attenuates the short-wavelength radiation from the system and re-emits it at longer wavelengths. This effect is clearly visible in Fig. \ref{fig:envgrida}. For $a=\qty{0.2}{au}$, the envelope density and optical depth are sufficiently large that the \qty{10}{\micro\meter} spectral feature is suppressed and the system SED is close to a blackbody spectrum. At smaller orbital distances, the temperature range in the system will also be reduced, resulting in narrower system SEDs.

Finally, systems at smaller orbital distances emit a lower total luminosity. In a given system, the energy budget is set not only by the depth of the potential well, but the value of the potential at the system boundary. For a fixed planet mass and mass accretion rate, material falling from a smaller Hill radius will therefore deliver less luminosity to the system. For large Hill radii, where $R_H\gg R_X$, these effects on the energy are minimal. However, for orbital radii $a\simeq\qty{0.2}{au}$ the Hill radius $R_H\simeq2R_X$ and the luminosity will be nearly halved in comparison to the fiducial value. The lower luminosity of the SEDs in Fig. \ref{fig:envgrida} reflect this fact.

{Fig. \ref{fig:PPDcomp} compares the SEDs of the planet/disk/envelope system to the region of the background protoplanetary disk that the circumplanetary system will replace. Comparing these spectra is important to understanding if the planet can be detected via contrast with the background disk. The temperature of this background disk is assumed to take the form}
\begin{equation}
    T_{\rm PPD}=T_{\rm X, PPD}\left(\frac{a}{R_{\rm X, PPD}}\right)^{-1/2}\,.
\end{equation}
{Here, $T_{\rm X, PPD}$ is the dust destruction temperature at \qty{1500}{K}. The radius $R_{\rm X, PPD}$ is the inner edge of the dust disk, which is assumed to occur at \qty{0.04}{au}, where the equilibrium temperature for a \qty{1.3}{L_\odot} star is the dust destruction temperature. Note that this is only an illustrative example, as protoplanetary disks can have a range of temperature profiles. In this specific case, the SED of a Hill sphere--sized region of the background disk will be }
\begin{equation}
    L_{\rm PPD,\nu}=4\pi\,\pi R_H^2\,B_\nu[T_{\rm PPD}(a)]\,.
\end{equation}
{The factor of $4\pi$ is a normalization convention for these plots and the factor of $\pi R_H^2$ is the area of the circumstellar disk that is replaced by the circumplanetary system. Since the luminosity of the background disk scales as $L\propto R_H^2T^4$, $R_H\propto a$, and $T\propto a^{-1/2}$, the background disk luminosity is independent of $a$ in this calculation. }

{For a fixed $M_p=\qty{1}{M_J}$ and $\dot{M}=\qty{10}{M_J/Myr}$, the flux of the protoplanet and the background disk are relatively equal in magnitude. However, the background disk spectrum is generally redder than the spectrum of the planetary system, especially at larger orbital radii. Forming systems will thus be easier to distinguish from the background disk at larger orbital separations. Note that this calculation does not account for attenuation of the planetary system by the background disk, which will likely have a significant impact on emergent radiative signatures. }

\begin{figure}
    \centering
    \includegraphics{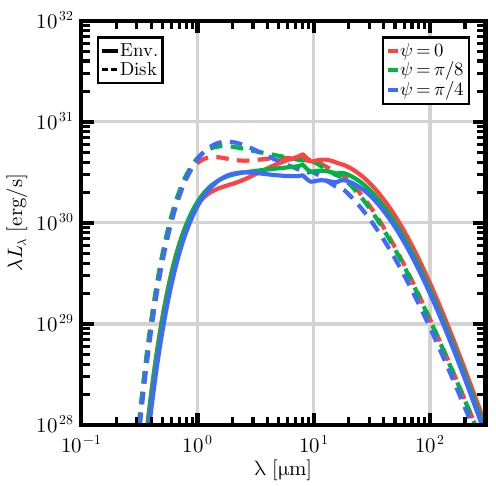}
    \caption{\textbf{Effects of Viewing Angle.} The spectral energy distributions of the planet/disk/envelope system for a range of viewing angles, where for $\psi=0$ the line of sight is aligned with the system pole and for $\psi=\pi/2$ the observer is in the equatorial plane. Aside from the viewing angle, the system parameters are set to their fiducial values (Table \ref{tab:canonvals}). The line style indicates whether the envelope has been included (solid) or not (dashed).}
    \vspace{-10pt}
    \label{fig:envgridpsi}
\end{figure}

\subsection{Viewing Angle}

Fig. \ref{fig:envgridpsi} shows how the system SEDs varies with the viewing geometry. For  spectra without an envelope (dashed lines), as the viewing geometry shifts towards the midlatitudes ($\psi\rightarrow\pi/4$), the line of sight encounters the disk at a more oblique angle. The observed solid angle of the disk is thus lower, which results in a dimmer disk spectrum ($\lambda\geq\qty{3}{\micro\meter}$). When the system is viewed near the equator ($\psi=\pi/2$, not shown), the observer will only be able to see the outer edge of the geometrically thick disks will strongly attenuate the planets' radiation and the SEDs will be almost entirely attenuated.

As the viewing angle approaches $\psi=\pi/4$, the inner edge of the disk near the planet becomes visible through the cavity at the disk center. {This results in a slight increase in the luminosity at around $\lambda=\qty{1}{\micro\meter}$, since the disk inner wall is nearly as hot as the planet itself. However, the temperature of the inner wall is somewhat overestimated, since \RAD does not account for dust sublimation. In any case, the increase in luminosity from the inner wall is minor. }

Finally, for the isotropic infall geometry used here, the envelope density is increased towards the equatorial plane. Observations at larger viewing angles will therefore encounter larger optical depths and the envelope will more significantly attenuate the radiation. In this case, however, the increasing envelope attenuation is insufficient to counteract the additional $\lambda\lesssim\qty{3}{\micro\meter}$ radiation from the hot inner disk, so this region is brighter for larger $\psi$ even if the envelope is included. Given the sensitivity of the SEDs to this parameter, more modeling is necessary to ensure the accuracy of the SEDs. 

However, it is generally expected that observations of circumplanetary disks will occur at small viewing angles, since conservation of angular momentum will require these disks to be somewhat aligned with the background circumstellar disk. If viewed edge-on, the circumstellar disk will completely absorb the radiative signatures of the circumplanetary system. Therefore, observations will necessarily be mostly face-on.

\begin{figure}
    \centering
    \includegraphics{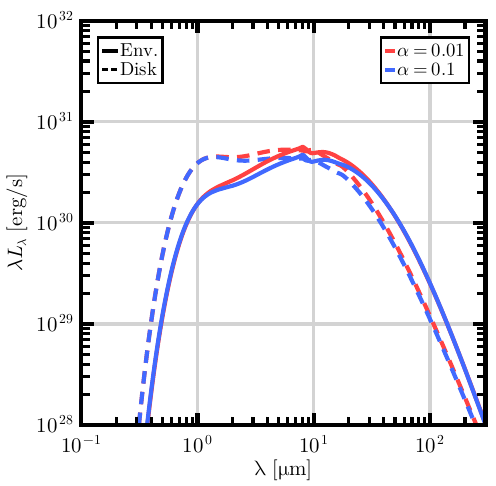}
    \caption{\textbf{Effects of Viscosity Parameter.} The spectral energy distributions of the planet/disk/envelope system for a range of viscosity parameters. The system is assumed to be viewed from above. Aside from the viscosity parameter, the system parameters are set to their fiducial values (Table \ref{tab:canonvals}). The line style indicates whether the envelope has been included (solid) or not (dashed).}
    \vspace{-10pt}
    \label{fig:envgridalpha}
\end{figure}

\subsection{Viscosity Parameter}

Varying the viscosity parameter $\alpha$ will dramatically affect the disk surface density and total mass but will only produce a small change in the system SEDs (see Fig. \ref{fig:envgridalpha}). For lower values of $\alpha$, the system is slightly brighter for $\lambda=\qtyrange{3}{30}{\micro\meter}$. This effect results from changing disk scale heights --- a lower viscosity means that the disk will have a larger surface density and therefore a larger total optical depth. The increased disk optical depth will naturally result in a hotter and geometrically thicker disk, which will absorb more of the planet's radiation and therefore must emit slightly more luminosity than the fiducial disk. This effect will be particularly noticeable in the inner disk, where the planet's radiation is the most significant and where the disk temperature has a blackbody peak in $\lambda=\qtyrange{3}{30}{\micro\meter}$. This region is precisely where the minor increase in the system SEDs are observed.

Note that the differences between these SEDs will only be minor when viewed along either the system pole or equator. If the system is viewed from the midlatitudes, the thick low-viscosity disks will heavily obscure both their central planets and the inner region of the disk. Since a thinner disk would only begin to obscure the planet at a larger viewing angle, the viscosity parameter can have significant qualitative effects on the disk SEDs when the system is viewed from these angles.

However, observations of circumplanetary disks are expected to occur at somewhat small viewing angles, where the effects of the viscosity parameter are minor. Constraints on this parameter therefore must rely on measurements of this parameter in the circumstellar disk and the assumption that the viscosity behaves similarly at these two distinct size scales. Fortunately, the insensitivity of the radiative signature to this parameter means that inaccuracies in this measurement are relatively unimportant to model fitting. 

{It is also possible that $\alpha$ will exhibit some radial dependence, which is not accounted for in these models. While a variable $\alpha$ would likely have a significant impact on the structure of the disk, the insensitivity of the SEDs to the value of $\alpha$ implies that varying this parameter across the disk would not impact the emergent radiative signatures. However, investigating the effect of variable $\alpha$ is beyond the scope of this paper.}

\begin{figure}
    \centering
    \includegraphics{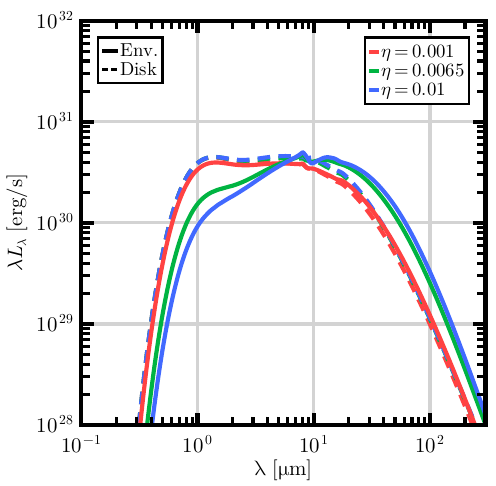}
    \caption{\textbf{Effects of Dust-to-Gas Ratio.} The spectral energy distributions of the planet/disk/envelope system for a range of dust-to-gas ratios. The system is assumed to be viewed from above. Aside from the dust-to-gas ratio, the system parameters are set to their fiducial values (Table \ref{tab:canonvals}). The line style indicates whether the envelope has been included (solid) or not (dashed).}
    \vspace{-10pt}
    \label{fig:envgrideta}
\end{figure}

\subsection{Dust-to-Gas Ratio}

The dust-to-gas ratio $\eta$ (variations shown in Fig. \ref{fig:envgrideta}) will have essentially no effect on the disk-only SEDs (dashed lines). After all, while a larger $\eta$ means a higher disk optical depth, these disks are already optically thick throughout. If the mass accretion rate were lower (e.g., $\dot{M}=\qty{1}{M_J\per\mega yr}$) and the outer disk were optically thin, then increasing the optical depth would cause some annuli to become optically thick. These annuli would be unable to radiate radially due to the increased optical depth, which would focus the radiation in the vertical direction and slightly increase the long-wavelength disk luminosity when viewed along the system pole. However, such an effect would be minor and only apply to disks with a lower mass accretion rate than the fiducial value. 

At this mass accretion rate, a far more significant effect is the increase in the optical depth of the envelope as the dust-to-gas ratio increases, which will absorb and reprocesses more of the planet and inner disk's short-wavelength radiation to longer wavelengths. The \qty{10}{\micro\meter} silicate absorption feature also becomes more prominent as the envelope optical depth increases. 

\begin{figure}
    \centering
    \vspace{-10pt}
    \includegraphics{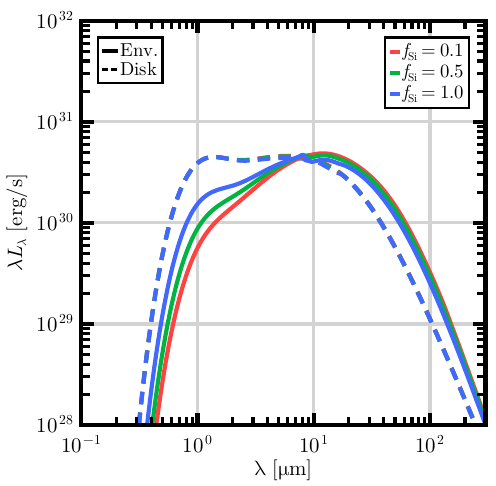}
    \caption{\textbf{Effects of Silicate Fraction.} The spectral energy distributions of the planet/disk/envelope system for a range of silicate fractions. The system is assumed to be viewed from above. Aside from the silicate fraction, the system parameters are set to their fiducial values (Table \ref{tab:canonvals}). The colors indicate the model used to calculate the spectrum, and the line style indicates whether the envelope has been included (dashed) or not (solid).}
    \vspace{-10pt}
    \label{fig:envgridsilfrac}
\end{figure}

\subsection{Silicate Fraction}

Fig. \ref{fig:envgridsilfrac} shows the changes in the SEDs induced by lowering the silicate fraction from $f_{\rm Si}=1$ to $f_{\rm Si}=0.1$ (while keeping $\eta$ constant). The dust composition has a very minor effect on the disk-only SEDs (dashed lines). Since the carbon opacity is higher than the silicate opacity (see Fig. \ref{fig:opacomp}), the disk optical depth is larger and the disk will be geometrically thicker and further concentrate radiation towards the system pole. Disks dominated by carbon opacity are thus brighter than those dominated by silicate opacity over the range $\lambda=\qtyrange{2}{20}{\micro\meter}$ at polar viewing angles. 

In addition, decreasing the silicate fraction increases the optical depth of the envelope, which will absorb more short-wavelength radiation and reprocess it to the infrared. Since the carbon opacity lacks the \qty{10}{\micro\meter} silicate feature, this feature also disappears from the spectrum once the carbon component dominates the dust opacity. Significantly, this change to the spectrum cannot be attributed to the luminosity or the optical depth of the envelope. Observing the absence or presence of the silicate feature will enable observations to directly constrain the silicate fraction and thus remove a free parameter from consideration. 

\begin{figure}
    \centering
    \vspace{-10pt}
    \includegraphics{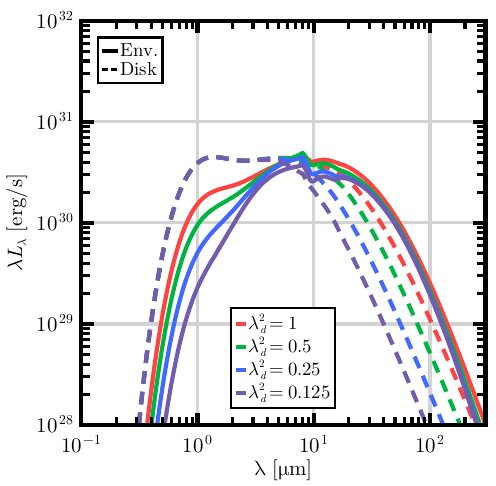}
    \caption{\textbf{Effects of Angular Momentum Deficit.} The spectral energy distributions of the planet/disk system if the angular momentum of the infalling material is reduced from the orbital momentum by a factor $\lambda_d$. This reduction will also lower the centrifugal radius of the disk by a factor $\lambda_d^2$. The system is assumed to be viewed from above. The line style indicates whether the envelope has been included (dashed) or not (solid).}
    \vspace{-10pt}
    \label{fig:envgridangmom}
\end{figure}

\subsection{Angular Momentum Deficit}

Recent work \citep{Adams2025} has expanded upon and generalized several of the assumptions made in our model. In particular, this new work includes more accurate infall trajectories, properly models the outer boundary condition of the disks, and accounts for angular momentum deficit in the infall. These corrections result in to smaller disk radii, higher envelope column densities, and higher disk accretion efficiencies. In the context of this work, the most significant effect will be the reduction of the outer disk radius that results from an angular momentum deficit. 

This model has generally assumed that the infalling material has an angular momentum equal to the planet's orbital angular momentum. The angular momentum of the incoming material will in general be reduced by some factor $\lambda_d$ and the centrifugal radius scales as $R_C\propto\lambda_d^{2}$. Fig. \ref{fig:envgridangmom} shows the effects of setting $\lambda_d^2=\numlist{1;0.5;0.25; 0.125}$, which is accounted for by removing the disk at $R>R_C$ (ensuring the total viscous heating luminosity is constant), and increasing the density of the envelope accordingly. The disk-only spectra (dashed lines) decrease at $\lambda\geq\qty{10}{\micro\meter}$ as the cold outer disk is removed. In this region, the spectra follow the Rayleigh-Jeans blackbody tail of the coldest effective temperature in the disk, which decreases as the disk shrinks. There are no changes in the disk-only spectra at wavelengths $\lambda\leq\qty{10}{\micro\meter}$. 

As the angular momentum deficit becomes more pronounced,
the disk radius shrinks, and incoming material falls farther inward before joining the disk. This change increases the density and optical depth of the envelope. The thicker envelope, in turn, increasingly absorbs the short-wavelength radiation from the planet (and inner disk) and re-emits this radiation at longer wavelengths, increasing the luminosity from \qtyrange{3}{100}{\micro\meter} relative to the disk-only spectrum. This reprocessed radiation essentially counteracts the loss of radiation at these wavelengths from the smaller disk. At the same time, the increased envelope optical depth makes the \qty{10}{\micro\meter} silicate feature more prominent.

Finally, as the angular momentum deficit increases (smaller values of $\lambda_d$), the spectral luminosity at the polar viewing angle decreases. However, the total luminosity of the system must remain constant by construction. This effect can be explained as follows ---as the disk shrinks, the outer edge of the disk will necessarily occur at a higher temperature. Since the disk is geometrically thick, the outer edge will primarily emit in the equatorial direction. As a result, smaller disks will emit a larger fraction of their luminosity in the equatorial plane, and emit a correspondingly smaller fraction of their luminosity towards the system pole. Larger angular momentum deficits will thus redirect radiation towards the system's equatorial plane, thereby lowering the luminosity at a polar viewing angle while maintaining the overall luminosity.

\section{Conclusion}\label{sec:disc}

\subsection{Summary}

In this paper, we have used a combination of the circumstellar disk model \DIAD and the radiative transfer model \RAD to self-consistently determine the structure of circumplanetary disks and their corresponding radiative signatures. 
These results show that the vertical temperature and density structure of these disks is inconsistent with a thin-disk approximation. Specifically, over most of parameter space, the circumplanetary disks have scale heights with aspect ratios in the range $H/R\sim0.1-0.25$ (Section \ref{sec:diskstruc}) {as have been found in prior work (e.g., \citealt{Szulagyi2016, Choksi2025})}. 
Models of the circumplanetary system must account for this geometry in order to accurately determine the structure of circumplanetary disks. These differences in the disk geometry will also have significant effects on the SEDs of the system. A comparison between the structure and radiative signatures of the \RADp model and a thin-disk semianalytic model is presented in Appendix \ref{sec:modelcomp}. 

While these disks could become gravitationally or thermally unstable in certain regions of parameter space, these disks are likely to be stable. The Toomre $Q$ parameter is $Q\sim\num{e3}$ for these disks (Section \ref{subsec:GI}), so viscosity or spiral shocks (e.g., \citealt{Zhu2016}) must be the dominant mechanism to transport material onto the protoplanet. As for the thermal instability, our models cannot find a steady-state solution at high mass accretion rates $\dot{M}\gtrsim\qty{60}{M_J\per\mega yr}$ (Section \ref{subsec:TI}). Although the value of this upper limit could be influenced by computational limitations,this cutoff is in close agreement with the mass accretion rate where the thermal instability is predicted to occur (Section \ref{subsec:TI}). In this regime, the system may experience variable accretion rates through the disk, analogous to FU Orionis accretion outbursts \citep{Hartmann1996, Hartmann2009}. While accretion signature variability has been observed on accreting protoplanets \citep{Bowler2025, Zhou2025, Close2025}, the mass accretion rates measured for these systems are far too low for the thermal instability to operate. In addition, it has recently been demonstrated that a cooling constraint on accreting envelopes \citep{Krapp2024} implies that the mass accretion rate cannot be larger than $\dot{M}\sim\qty{70}{M_J\per\mega yr}$ \citep{Adams2025}. Since this value is close to the mass accretion rate where the thermal instability sets in, thermally unstable circumplanetary disks are unlikely.

Although the circumplanetary envelope can be optically thick, additional heating of the disk by the envelope is unimportant (Section \ref{sec:envheat}). Heating of the disk by the envelope provides at most $10\%$ of the disk temperature and only reaches these levels for high mass accretion rates when the disk is embedded in an optically thick envelope that obscures the relevant region. Although the temperature increases modestly, the density structure is largely unaffected. 

Section \ref{sec:disksig} shows how changing system parameters affect the radiative signatures of the system. Although the product $M_p\dot{M}$ characterizes the luminosity, these system parameters can be constrained individually by the spectral color. Specifically, increasing the mass accretion rate results in redder spectra while increasing the planet mass leads to bluer spectra. As the mass accretion rate increases, the disk scale height and the envelope optical depth increase, which reprocesses more radiation to longer wavelengths. On the other hand, a higher planet mass will reduce both the disk scale height and the envelope optical depth and produce the opposite effect.

{Section \ref{sec:disksig} also shows that the radiative signatures at $\lambda\geq\qty{10}{\micro\meter}$ are independent of the infall geometry to first order. Observations in the near infrared will thus be necessary to determine the geometry of the accretion flow. In addition, since lowering the planet's orbital distance from the star leads to narrower spectral energy distributions, the location of the planet has a significant impact on its radiative signatures. By a similar mechanism, increasing the angular momentum deficit leads to narrower SEDs and a stronger \qty{10}{\micro\meter} opacity feature. The relative strength of this feature can thus be used in tandem with a measured angular separation to measure the angular momentum of the infalling material. }

Given these variations, future observations in the optical and the near-infrared will be able to constrain the parameters of these systems. Since the differences in the SEDs cover broad wavelength bands, low-resolution spectra across a wide range of wavelengths, or even multi-wavelength photometry, will likely be most effective at constraining system parameters. 

\subsection{Discussion}

Observations of accreting protoplanetary systems are necessary to constrain their accretion behavior and the evolution of giant planets. Fortunately, with the dawn of \textit{JWST} and the forthcoming ELT, such observations are expected in the near future. A brief discussion of these observations is warranted. 

Currently, observations of young accreting protoplanets are generally in the optical, the near infrared (NIR), and the submillimeter; detections of protoplanet accretion to date have been dominated by the optical \qty{0.656}{\micro\meter} H$\alpha$ line (e.g., \citealt{Ringqvist2023, Viswanath2024}). Additional NIR hydrogen emission lines have provided supporting evidence for this accretion, but are generally weaker than the H$\alpha$ line. Most critically, detection of spatially-resolved dust emission at \qty{855}{\micro\meter} has provided strong evidence of a protoplanetary disk around PDS 70 b/c \citep{Isella2019,Benisty2021}. Due to the low mass accretion rates of these systems ($\dot{M}\simeq\qtyrange{0.01}{0.1}{M_J/\mega yr}$, \citealt{Aoyama2019, Haffert2019, Thanathibodee2019, Hashimoto2020}), this detection likely represents an evolved circumplanetary disk, which is further supported by radio observations \citep{Dominguez-Jamett2025}. 

{While there have been several claimed protoplanet detections, the implied mass accretion rates are all small ($\dot{M}\lesssim\qty{0.1}{M_J/Myr}$; \citealt{Aoyama2019, Li2025, Close2025a, Cugno2019}). At these low mass accretion rates, the internal luminosity of the protoplanet may be significant relative to the energy delivered by accretion. Our models thus require further modification to be applicable to these systems. In addition, radio, optical, and infrared surveys of circumstellar disk gaps \citep{Zurlo2020, Asensio-Torres2021, Andrews2021, Wallack2024a} have failed to detect accreting protoplanets. While the upper limits on the mass accretion rates are not stringent (generally $\dot{M}\lesssim\qty{10}{M_J/Myr}$), the low accretion rates of known protoplanets and the nondetection of protoplanets in these gaps implies that large accretion rates are rare once a planet is no longer embedded in the circumstellar disk. }

While the emission of circumplanetary disks is expected to peak in the NIR, it is difficult to spatially disentangle these observations from the background circumstellar disk. For large-scale surveys, it will likely be necessary to focus on spatially-resolved H$\alpha$ or submillimeter emission as a tracer of protoplanet accretion or a dusty circumplanetary disk. These disks may also be detected in the radio \citep{Zhu2015, Zhu2018, Dominguez-Jamett2025}. Once detected, a subset of objects could be targeted for follow-up observations of the circumplanetary disk in the infrared or visible to further constrain the system properties. Since the SEDs are relatively constant across NIR bands for many system parameters, low-resolution spectra across a wide range of wavelengths or multi-wavelength photometry will likely be especially effective at constraining system parameters. 

The distinctions in the SEDs can be primarily ascribed to changes in the luminosity ($M_p$, $\dot{M}$) and the envelope optical depth. Most of the parameters considered here only modify the radiative signatures by modifying these two intermediate parameters, meaning that there is some degeneracy in the system radiative signatures. There are few exceptions, primarily where parameters act by changing the geometry of the disk, which may help in breaking this degeneracy. Further work is necessary to precisely determine the parameters that observations will be able to constrain. 

Unfortunately, circumplanetary disks will not be observable during the entirety of the runaway accretion phase. When runaway accretion starts, the system is completely embedded in the circumstellar disk and accretes from the ambient material, which strongly attenuates the SED (see Fig. 15 in \citealt{Taylor2025}). However, once the planet clears a gap, the ambient material is removed and the mass accretion rate will naturally drop off. Since the circumplanetary disk viscosity will generally be sufficient to maintain a disk in steady-state with the incoming mass accretion, systems in this mode of embedded accretion will be difficult to detect. At somewhat later stages, however, a concentrated stream of material across the gap can provide a significant mass flux and allow the system radiation to escape and become observable. In this later stage, the structure and radiative signatures of the circumplanetary system can be reasonably modeled as a disk and planet alone, without the surrounding envelope. SEDs for this architecture are shown in Figs. \ref{fig:envgridMpd}--\ref{fig:envgridangmom} (dashed lines). 

Even when accretion takes place along streamers, observations can constrain the properties of these circumplanetary disks, since the SEDs still depend strongly on several parameters (e.g., planet mass, mass accretion rate, viewing angle, semimajor axis, angular momentum deficit). Observations of these systems will thus be able to constrain some of their properties. Fortunately, while the opacity is a significant unknown in these models, the disks are generally optically thick and do not display strong features. The SEDs of a streamer-fed system are thus largely independent of the opacity.

{It is also worthwhile to compare these results to those of previous authors. \citet{Choksi2025} calculated the structure of the circumplanetary disk by assuming that the disk was in a viscous steady-state and also found that these disks will be geometrically thick, with similar aspect ratios. In the limited region of parameter space where our models overlap, our SEDs are broadly similar. Other work simulating observations based on hydrodynamical simulations \citep{Szulagyi2018, Szulagyi2019, Chen2022} predict that the circumplanetary region will significantly attenuate the planet's radiation and render it undetectable. This is in contrast to our results, where the planet is generally not significantly attenuated by the circumplanetary disk, although the circumplanetary envelope can significantly absorb the planet's radiation. This difference likely arises from different choices for the inner boundary condition, in particular the magnetic truncation radius.}

{Throughout this work, we have assumed that the circumplanetary disk and envelope are azimuthally symmetric. This assumption is justified for the circumplanetary disk, since differential rotation will act to smooth out any azimuthal asymmetries that are not being continually driven. Since the timescale of the disks' orbit (\qty{1}{yr}) is much less than the infall timescale (\qty{1}{Myr}), the disk will have ample time to correct any asymmetry. While the envelope may have azimuthal asymmetries driven by spiral density waves and/or a ``headwind'' from sub-Keplerian rotation in the background circumstellar disk \citep{Cimerman2017, Krapp2022, Kuwahara2024}, these effects are expected to be subdominant. Since these envelopes are generally optically thin to their own radiation, the temperature distribution of the envelope is not strongly affected by such an asymmetry. Finally, observations will be unable to resolve the circumplanetary system and will capture the entire system in a single pixel. The emergent SEDs will therefore be averaged over the entire system, washing out any azimuthal perturbation.}

This work has made several assumptions that can be generalized. First, the model assumes that the dust has a maximum size of \qty{100}{\micro\meter}. In the background circumstellar disk, the large dust grains will settle to the midplane, while small dust grains are well-coupled to the gas and can remain in the upper levels in the disk. Over time, dust in the midplane is expected to grow larger and eventually produce planetesimals. Since the planetary systems considered here will be accreting from the midplane of relatively old circumstellar disks (where most of the disk mass is), the circumplanetary environment will be preferentially enriched in dust at this size scale. {However, given the temperatures of these circumplanetary disks, it is likely that \qty{100}{\micro\meter} dust will be destroyed in collisions. Future work incorporating dust size evolution calculations would be worthwhile to better understand the growth of dust in these systems.}

In addition, this model relied on the \DIAD model to construct the disks, which assumes that the disk has a constant mass accretion rate throughout. In reality, the mass accretion rate will vary significantly and change sign at some point in the disk, which may have significant effects on the structure of the disk and affect the resulting radiative signatures. In future work, we will account for this variable $\dot{M}$. {Similarly, the viscosity parameter $\alpha$ is assumed to be fixed throughout the disk. In reality, it is likely that $\alpha$ has some radial dependence, the investigation of which is also left for future work.}

{This calculation has also treated the scattering of light by simply adding the scattering opacity to the absorption opacity. However, accurately modeling the scattering has a negligible effect on the disk structure or the radiative signatures presented here. The scattering of the radiation from the planet is only relevant when the system is viewed from near the equatorial plane. In this geometry, the circumstellar disk will strongly attenuate the system's radiation and the circumplanetary system will not be detectable.}

Finally, we have neglected line emission and absorption in these models. To date, few spectral observations of circumplanetary disks exist, a situation that is unlikely to change in the immediate future. We have therefore concentrated on characterizing the underlying continuum, which will dominate the radiative signatures. As observations advance and circumplanetary disks are detected and characterized, line modeling will become important and will be addressed in future work.

As observations of circumplanetary disks become available in the near future, efficient and self-consistent physical models will be necessary to model the mechanics of these systems. This work models the structure of circumplanetary disks and shows that their spectra are complex functions of the system parameters, further advancing the goal of understanding the physics of giant planet formation. 

\section*{Acknowledgments}

A.G.T. acknowledges support from the Fannie and John Hertz Foundation and the University of Michigan's Rackham Merit Fellowship Program. F.C.A. is supported in part by NSF Grant No. 2508843 and by the Leinweber Institute for Theoretical Physics at the University of Michigan. 
This paper made use of the Julia programming language \citep{Julia} and the plotting package Makie \citep{Makie}.

\appendix

\section{{Model Convergence}}\label{app:conv}

\begin{figure}
    \centering
    \vspace{-10pt}
    \includegraphics{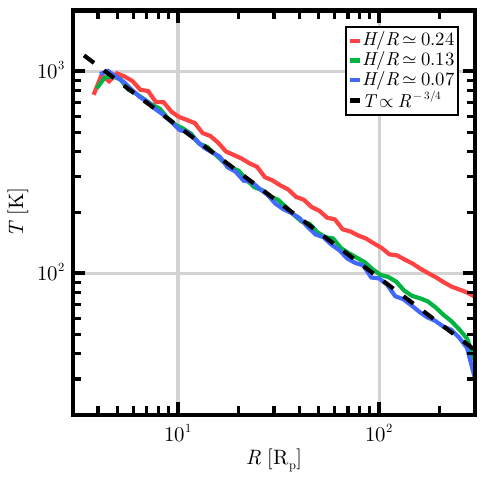}
    \caption{{\textbf{Convergence of the Disk Temperature Distribution.} The photosphere temperature distribution versus radius for a series of test disk models. The mass accretion rates are set to $\dot{M}=\qtylist{10;3.16;1}{M_J/Myr}$ and the planet masses are set to $M_p=\qtylist{1;3.16;10}{M_J}$ respectively, lowering the disk aspect ratios ($H/R=\numlist{0.24;0.13;0.07}$) while keeping the luminosity fixed. The inner radius of the disk is set to $R_X=\qty{3.8}{R_p}$ and the outer radius is set to $R_C=\qty{300}{R_p}$. The temperature distribution expected for a thin disk ($T\propto R^{-3/4}$) is also shown. }}
    \vspace{-10pt}
    \label{fig:temp_noplanet}
\end{figure}

{This section demonstrates that the \RADp model correctly converges to the simple thin $\alpha$-disk model as the disk scale height is reduced. We present a series of models where the luminosity of the planet is removed so that the disk is solely heated by the viscous accretion flow. In this limit, the effective temperature of the disk should take the form $T\propto R^{-3/4}$ and the SED should take the form $\nu L_\nu\propto\nu^{4/3}$. This result relies on the assumption that all of the energy released by the accretion in a given annulus is confined to that annulus --- that is, that energy is not transferred radially. For geometrically thick disks such as those found in this work, this assumption does not hold. For thin disks, the \RADp model will reproduce the expected disk form.}

{The scale height of the disk will increase with the mass accretion rate and decrease for increasing planet mass. Here we calculate the temperature distribution and SEDs for models with $\dot{M}=\qtylist{10;3.16;1}{M_J/Myr}$ and $M_p=\qtylist{1;3.16;10}{M_J}$, respectively. These models have a fixed luminosity (since $L\propto M_p\dot{M}$) but varying scale heights. To ensure that these models can be compared, the outside edge of the disk is fixed at $R_C=\qty{300}{R_p}$ and the inside edge is fixed at $R_X=\qty{3.8}{R_p}$. The aspect ratios of the disks are $H/R=\numlist{0.24;0.13;0.07}$.}

\begin{figure}
    \centering
    \vspace{-10pt}
    \includegraphics{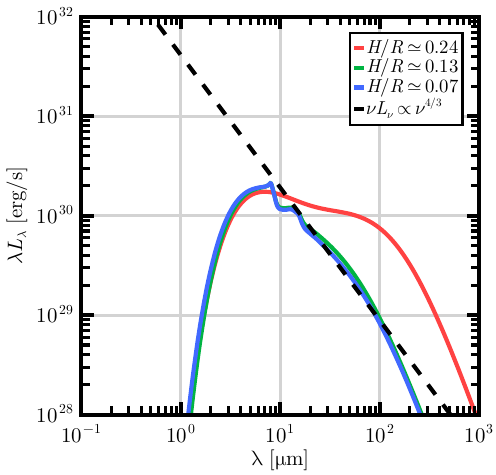}
    \caption{{\textbf{Convergence of the Disk SEDs.} The emergent SEDs for a series of test disk models. The mass accretion rates are set to $\dot{M}=\qtylist{10;3.16;1}{M_J/Myr}$ and the planet masses are set to $M_p=\qtylist{1;3.16;10}{M_J}$ respectively, lowering the disk aspect ratios ($H/R=\numlist{0.24;0.13;0.07}$) while keeping the luminosity fixed. The inner radius of the disk is set to $R_X=\qty{3.8}{R_p}$ and the outer radius is set to $R_C=\qty{300}{R_p}$. The expected power-law SED ($\nu L_\nu\propto\nu^{4/3}$) is also shown. }}
    \vspace{-10pt}
    \label{fig:sed_noplanet}
\end{figure}

{Fig. \ref{fig:temp_noplanet} shows the temperature at the disk photosphere versus radius for these models, along with the expected $T\propto R^{-3/4}$ power law. As the disk aspect ratio is reduced, the temperature distribution clearly approaches the expected functional form. At the same time, the SEDs of these models (shown in Fig. \ref{fig:sed_noplanet}) also approach the expected distribution as the disk becomes geometrically thin, albeit modulated by the opacity features as seen in prior work (Fig. 13 in \citealt{DAlessio1998} and Fig. 6 in \citealt{Zhu2007}). This convergence demonstrates the stability of the \RADp model. Finally, we note that the degree to which these models fail to match the approximation is the central subject of this work. }

\section{Shock Luminosity}\label{app:shocks}

In this section, we discuss the effects of the planetary accretion shock on the circumplanetary disk. {In the main body of this work, the energy delivered by accretion is uniformly spread across the entire surface of the planet. However, in reality magnetic accretion will deliver material to a fraction of the planet's surface, which will be hotter than the rest of the planet. Some fraction of the energy delivered to the planet's surface will be emitted by this ``hot spot'' while the remaining energy is emitted by the planet's entire surface area. In addition to the hot spot efficiency, the hot spot is characterized by a temperature $T_{\rm hs}$. For the sake of definiteness, here $T_{\rm s}\simeq\qty{8000}{K}$. Since this value is likely hotter than the typical shock temperature on an accreting giant planet (being appropriate for T Tauri stars, e.g., \citealt{Calvet1998}), the results of this section serve as an upper limit for the effects of a shock on the spectra.}

In general, the accretion delivers a total luminosity $L_p$ to the planet, which can be divided into two portions --- the hot spot luminosity $L_{\rm hs}$ and the planet's thermal emission $L_{\rm th}$. The relationship between these quantities is parameterized by the efficiency so that $\epsilon\equiv L_{\rm hs}/L_p$. The remaining luminosity is therefore released by the planet's thermal emission at a temperature
\begin{equation}
    T_p=\left(\frac{(1-\epsilon)L_p}{4\pi \sigma R_p^2}\right)^{1/4}\,.
\end{equation}
With $L_p$ given by Eq. \eqref{eq:Lp} and the shock temperature fixed at $T_{\rm hs}=\qty{8000}{K}$, the shock and thermal emission of the planet are completely specified for an efficiency $\epsilon$. In particular, the total emission from the planet's surface at a frequency $\nu$ is given by 
\begin{equation}
    L_\nu=\frac{\epsilon L_p}{\sigma T_s^4}\pi B_\nu(T_s)+4\pi^2R_p^2S_{\nu}(T_p)\,,
\end{equation}
where $S_{\nu}$ is the \texttt{SONORA Bobcat} spectrum for the corresponding temperature. Note that in this setup, energy is conserved by construction.

\begin{figure}
    \centering
    \includegraphics{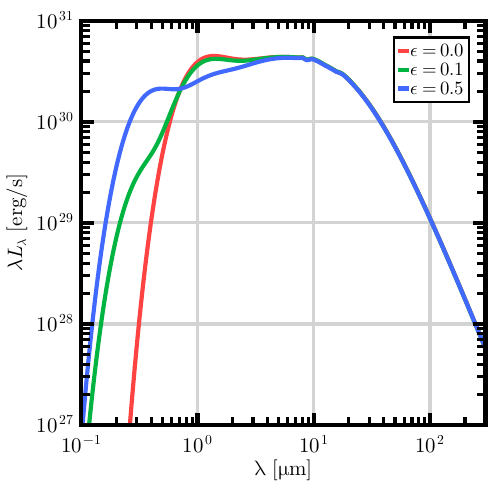}
    \caption{\textbf{Shock Radiative Signatures.} The radiative signatures predicted by the \RAD model when some of the luminosity is emitted in the shock. The system is viewed from above through an isotropic-infall envelope. All parameters are set to fiducial values.}
    \label{fig:shockspec}
\end{figure}

Throughout this work, we have assumed that $\epsilon=0$, so that the entirety of the accreted energy is released by the thermal emission of the planet. Fig. \ref{fig:shockspec} shows the SED predicted by \RAD for $\epsilon=0.1$ and $\epsilon=0.5$, assuming isotropic infall and a polar viewing angle. The primary effect is the reduction of the planet's thermal radiation at \qtyrange{0.7}{2}{\micro\meter} in favor of the shock radiation in the optical ($\lambda\leq\qty{0.7}{\micro\meter}$). The spectrum from the disk is essentially unchanged for $\lambda\gtrsim\qty{3}{\micro\meter}$, and the disk structure is similarly unmodified. Therefore, while shocks have little effect on the disk structure, they may be important in constraining the planet's SED. Joint observations in the optical and NIR would be necessary to constrain the importance of shocks during these stages of planet formation. 

\begin{figure}
    \centering
    \includegraphics{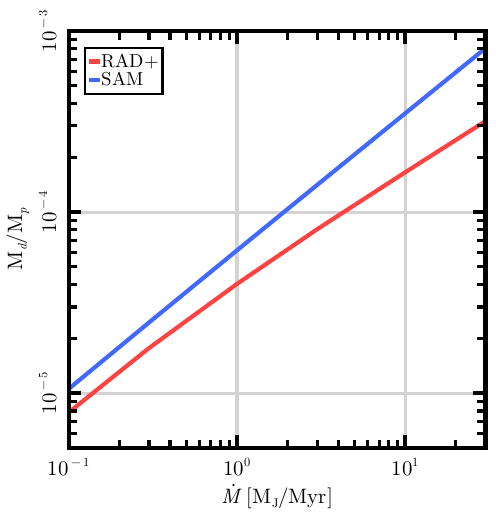}
    \caption{\textbf{Disk/Planet Mass Ratio.} The ratio between the disk mass $M_d$ and the planet mass $M_d$ versus mass accretion rate $\dot{M}$. All other parameters are set to their fiducial values. The viscosity parameter is set to $\alpha=0.1$ here, but the disk mass will have the form $M_d\propto\alpha^{-1}$.  }
    \label{fig:diskmass}
\end{figure}

\section{Model Comparison}\label{sec:modelcomp}

This section compares the structure and radiative signatures of the circumplanetary disks calculated by our numerical \RADp model to the semianalytic model (henceforth SAM) introduced in previous work. 

\subsection{Semianalytic Model}\label{subsec:TA25}

First, the SAM model must be described. This approach assumed that the planet is a blackbody, infall is ballistic, and that each point on the geometrically-thin circumplanetary disk emits as a blackbody at the corresponding disk temperature. This subsection briefly reviews the structure of this model. 

The planet resides at the center of the system and is surrounded by its disk and the infalling envelope. The planet is assumed to be a blackbody with luminosity and temperature given by Eqs. \eqref{eq:Lp} and \eqref{eq:Tp}.  

The circumplanetary disk extends from the magnetic truncation radius $R_X$ to the centrifugal radius $R_C$. The disk is assumed to be geometrically thin, with a temperature distribution $T_d(r)\propto r^{-3/4}$. The disk luminosity is provided by the steady-state mass accretion with no contribution from the radiation of the central planet, and each disk annulus emits a blackbody spectrum at its local temperature. The temperature is normalized so that the disk as a whole emits a luminosity $L_d=f_d GM_p\dot{M}/2R_X$, where $f_d$ is the fraction of material that is accreted directly onto the disk and $R_X$ is the magnetic truncation radius. 

For the purposes of calculating an SED, the temperature distribution of the infalling envelope is assumed to be spherically symmetric, in thermal equilibrium, and optically thin to its own radiation, so that the planet and disk luminosity the envelope absorbs is equal to the luminosity it emits. The opacity of the envelope is set to be a power law opacity, so that $\kappa_\nu=\kappa_0(\nu/\nu_0)^\gamma$. In order to accurately match the \RADp dust opacity, the power-law index $\gamma=1/2$, $\kappa_0=\qty{3}{cm^2/g}$ and $\nu_0=\qty{3e13}{Hz}$. Except for the silicate feature at \qty{10}{\micro\meter}, the silicate opacity is relatively smooth over the relevant wavelength regime and is in good agreement with the $\gamma=1/2$ power law (see Fig. \ref{fig:opacomp}).  

The radiative contributions of the planet, disk, and envelope are calculated separately. The planet emits a SED with a blackbody spectrum that is attenuated by the envelope along a ray extending from the center of the planet to the observer. For each point on the disk surface, the emission in a given direction is also attenuated by the envelope. The total disk emission in a given direction is then calculated by integrating over the disk surface, accounting for the attenuation along each corresponding ray through the envelope. Finally, emission of the envelope is integrated across the entire system without attenuation, since we assumed that the envelope is optically thin to its own radiation.

\subsection{Disk Structure}

A major difference between the SAM and \RADp models is the structure of the circumplanetary disk. These differences are primarily driven by SAM's assumption that the disk is geometrically thin, which is not the case in the \RADp model. There are a few additional points of note. 

For one, \RADp predicts that each annulus of the circumplanetary disk will be hotter than the emission temperature calculated by SAM. This difference results from the fact that \RADp includes the protoplanet's radiation when calculating the disk model, which will be absorbed by the disk and necessarily increase its temperature. In addition, since the \RADp disks are generally optically thick, the midplane temperatures of these disks will be hotter than the emission temperature calculated by energy conservation arguments alone. As discussed in Section \ref{sec:diskstruc}, the disk midplane to be hotter than the upper layers, since a steep temperature gradient is necessary to transport the viscous heating luminosity to the emission surface of the disk. 

The hotter temperatures of the \RADp disks also mean that the \RADp disks are less massive than their corresponding SAM disks. For a fixed value of $\alpha$, higher temperatures necessarily imply a higher viscosity. Since the total energy emitted will be approximately conserved, a hotter disk will therefore have a smaller surface density. As a corollary, SAM predicts larger disk masses than \RADp. Fig. \ref{fig:diskmass} shows the disk-planet mass ratio $M_d/M_p$ versus mass accretion rate. For both models, the total disk mass is found by integrating the surface density across the disk. As discussed, SAM calculates that a given disk is approximately twice as massive as the corresponding \RADp disk. While the maximum mass ratio of these disks are at the relatively low value of $M_d/M_p\sim\num{e-3}$, this value is calculated assuming a viscous $\alpha$ parameter of \num{0.1}. For smaller viscosities more typical of circumstellar disks (i.e., $\alpha\sim\num{e-4}$), the disk mass will be comparable to the planet mass.

\begin{figure}
    \centering
    \vspace{-10pt}
    \includegraphics{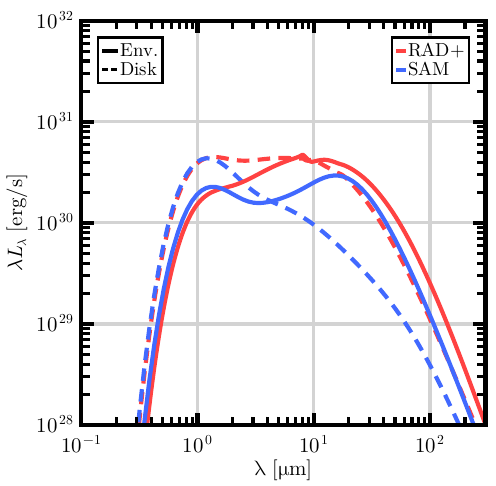}
    \caption{\textbf{SED Model Comparison.} The system radiative signatures predicted by \RADp (red) and SAM (blue). The system parameters are set to their fiducial values (Table \ref{tab:canonvals}). The colors indicate the model used to calculate the spectrum, and the line style indicates whether the envelope has been included (dashed) or not (solid).}
    \vspace{-10pt}
    \label{fig:modelcomp}
\end{figure}

\subsection{Radiative Signatures}

Fig. \ref{fig:modelcomp} shows the SEDs predicted by \RADp and SAM for the fiducial model, both including and excluding the envelope. Unfortunately, the SAM model is a relatively poor approximation of the \RADp model, so incorporating the disk's geometry will be necessary to model observations. 

For the disk-only spectra (dashed lines), the two models generally agree on the planet spectrum for $\lambda\leq\qty{1}{\micro\meter}$. However, \RADp predicts a significantly brighter disk than SAM does. This difference is a natural consequence of the relatively thick disks predicted by \RADp. Unlike a thin disk, the surface of a thick disk is blocked by the rest of the disk when viewed from near the equatorial plane. Therefore, thick disks will necessarily concentrate their radiation towards the system pole, increasing the luminosity in this direction. The disk geometry is primarily responsible for the brighter \RADp spectrum, which is a factor of four brighter than the SAM disks. 

The differences between these models will be even more significant when the system is viewed from near the equatorial plane. In this observation geometry, the thick \RADp disks will strongly attenuate the planets' radiation and the SEDs will be significantly dimmer. However, the surface of the thin disk will still be visible and contribute significant luminosity. The viewing-angle-integrated luminosity of these systems is approximately equal, although numerical noise and slightly different disk models introduce discrepancies in the luminosity of {a few percent}. 

The spectra are even more distinct when accounting for the attenuation of the envelope. First, the \RADp spectrum exhibits the \qty{10}{\micro\meter} silicate feature, which is lacking in the SAM spectrum. More critically, the SAM spectrum has two humps, with a peak at $\lambda\simeq\qty{1}{\micro\meter}$ and one at $\lambda=\qty{20}{\micro\meter}$. The first peak is due to the planet spectrum, which is attenuated heavily at shorter wavelengths and drops off at longer wavelengths. The second peak is due to the envelope, which has a somewhat narrow range of temperatures and so emits in a broad hump. In comparison, the optical depth of the envelope to its own radiation is nonzero in the \RADp model, which necessitates a stronger temperature gradient and thereby spreads out this hump. These differences mean that models of the circumplanetary environment will likely have to account for the envelope's self-attenuation (as well as the disk geometry) to accurately predict the radiative signatures of these systems. 

\bibliographystyle{aasjournal}
\bibliography{main.bib}

@article{Cugno2019,
	title = {A search for accreting young companions embedded in circumstellar disks. {High}-contrast {Hα} imaging with {VLT}/{SPHERE}},
	volume = {622},
	issn = {0004-6361},
	url = {https://ui.adsabs.harvard.edu/abs/2019A&A...622A.156C},
	doi = {10.1051/0004-6361/201834170},
	urldate = {2025-10-28},
	journal = {Astronomy and Astrophysics},
	author = {Cugno, G. and Quanz, S. P. and Hunziker, S. and Stolker, T. and Schmid, H. M. and Avenhaus, H. and Baudoz, P. and Bohn, A. J. and Bonnefoy, M. and Buenzli, E. and Chauvin, G. and Cheetham, A. and Desidera, S. and Dominik, C. and Feautrier, P. and Feldt, M. and Ginski, C. and Girard, J. H. and Gratton, R. and Hagelberg, J. and Hugot, E. and Janson, M. and Lagrange, A.-M. and Langlois, M. and Magnard, Y. and Maire, A.-L. and Menard, F. and Meyer, M. and Milli, J. and Mordasini, C. and Pinte, C. and Pragt, J. and Roelfsema, R. and Rigal, F. and Szulágyi, J. and van Boekel, R. and van der Plas, G. and Vigan, A. and Wahhaj, Z. and Zurlo, A.},
	month = feb,
	year = {2019},
	note = {Publisher: EDP
ADS Bibcode: 2019A\&A...622A.156C},
	pages = {A156},
}

@article{Asensio-Torres2021,
	title = {Perturbers: {SPHERE} detection limits to planetary-mass companions in protoplanetary disks},
	volume = {652},
	issn = {0004-6361},
	shorttitle = {Perturbers},
	url = {https://ui.adsabs.harvard.edu/abs/2021A&A...652A.101A},
	doi = {10.1051/0004-6361/202140325},
	urldate = {2025-10-28},
	journal = {Astronomy and Astrophysics},
	author = {Asensio-Torres, R. and Henning, Th. and Cantalloube, F. and Pinilla, P. and Mesa, D. and Garufi, A. and Jorquera, S. and Gratton, R. and Chauvin, G. and Szulágyi, J. and van Boekel, R. and Dong, R. and Marleau, G.-D. and Benisty, M. and Villenave, M. and Bergez-Casalou, C. and Desgrange, C. and Janson, M. and Keppler, M. and Langlois, M. and Ménard, F. and Rickman, E. and Stolker, T. and Feldt, M. and Fusco, T. and Gluck, L. and Pavlov, A. and Ramos, J.},
	month = aug,
	year = {2021},
	note = {Publisher: EDP
ADS Bibcode: 2021A\&A...652A.101A},
	pages = {A101},
}

@article{Andrews2021,
	title = {Limits on {Millimeter} {Continuum} {Emission} from {Circumplanetary} {Material} in the {DSHARP} {Disks}},
	volume = {916},
	issn = {0004-637X},
	url = {https://ui.adsabs.harvard.edu/abs/2021ApJ...916...51A},
	doi = {10.3847/1538-4357/ac00b9},
	urldate = {2025-10-28},
	journal = {The Astrophysical Journal},
	author = {Andrews, Sean M. and Elder, William and Zhang, Shangjia and Huang, Jane and Benisty, Myriam and Kurtovic, Nicolás T. and Wilner, David J. and Zhu, Zhaohuan and Carpenter, John M. and Pérez, Laura M. and Teague, Richard and Isella, Andrea and Ricci, Luca},
	month = jul,
	year = {2021},
	note = {Publisher: IOP
ADS Bibcode: 2021ApJ...916...51A},
	pages = {51},
}

@article{Wallack2024a,
	title = {A {Survey} of {Protoplanetary} {Disks} {Using} the {Keck}/{NIRC2} {Vortex} {Coronagraph}},
	volume = {168},
	issn = {0004-6256},
	url = {https://ui.adsabs.harvard.edu/abs/2024AJ....168...78W},
	doi = {10.3847/1538-3881/ad390c},
	urldate = {2025-10-28},
	journal = {The Astronomical Journal},
	author = {Wallack, Nicole L. and Ruffio, Jean-Baptiste and Ruane, Garreth and Ren, Bin B. and Xuan, Jerry W. and Villenave, Marion and Mawet, Dimitri and Stapelfeldt, Karl and Wang, Jason J. and Liu, Michael C. and Absil, Olivier and Alvarez, Carlos and Bae, Jaehan and Bond, Charlotte and Bottom, Michael and Calvin, Benjamin and Choquet, Élodie and Christiaens, Valentin and Cook, Therese and Femenía Castellá, Bruno and Gomez Gonzalez, Carlos and Guidi, Greta and Huby, Elsa and Kastner, Joel and Knutson, Heather A. and Meshkat, Tiffany and Ngo, Henry and Ragland, Sam and Reggiani, Maddalena and Ricci, Luca and Serabyn, Eugene and Uyama, Taichi and Williams, Jonathan P. and Wizinowich, Peter and Zawol, Zoe and Zhang, Shangjia and Zhu, Zhaohuan},
	month = aug,
	year = {2024},
	note = {Publisher: IOP
ADS Bibcode: 2024AJ....168...78W},
	pages = {78},
}

@article{Zurlo2020,
	title = {The widest {Hα} survey of accreting protoplanets around nearby transition disks},
	volume = {633},
	issn = {0004-6361},
	url = {https://ui.adsabs.harvard.edu/abs/2020A&A...633A.119Z},
	doi = {10.1051/0004-6361/201936891},
	urldate = {2025-10-28},
	journal = {Astronomy and Astrophysics},
	author = {Zurlo, A. and Cugno, G. and Montesinos, M. and Perez, S. and Canovas, H. and Casassus, S. and Christiaens, V. and Cieza, L. and Huelamo, N.},
	month = jan,
	year = {2020},
	note = {Publisher: EDP
ADS Bibcode: 2020A\&A...633A.119Z},
	pages = {A119},
}

@article{Cimerman2017,
	title = {Hydrodynamics of embedded planets’ first atmospheres – {III}. {The} role of radiation transport for super-{Earth} planets},
	volume = {471},
	issn = {0035-8711},
	url = {https://doi.org/10.1093/mnras/stx1924},
	doi = {10.1093/mnras/stx1924},
	number = {4},
	urldate = {2025-10-28},
	journal = {Monthly Notices of the Royal Astronomical Society},
	author = {Cimerman, Nicolas P. and Kuiper, Rolf and Ormel, Chris W.},
	month = nov,
	year = {2017},
	pages = {4662--4676},
}

@article{Kuwahara2024,
	title = {Analytic description of the gas flow around planets embedded in protoplanetary disks},
	volume = {682},
	issn = {0004-6361},
	url = {https://ui.adsabs.harvard.edu/abs/2024A&A...682A..14K},
	doi = {10.1051/0004-6361/202347530},
	urldate = {2025-10-28},
	journal = {Astronomy and Astrophysics},
	author = {Kuwahara, Ayumu and Kurokawa, Hiroyuki},
	month = feb,
	year = {2024},
	note = {Publisher: EDP
ADS Bibcode: 2024A\&A...682A..14K},
	pages = {A14},
}

@article{Krapp2022,
	title = {The {3D} {Dust} and {Opacity} {Distribution} of {Protoplanets} in {Multifluid} {Global} {Simulations}},
	volume = {928},
	issn = {0004-637X},
	url = {https://ui.adsabs.harvard.edu/abs/2022ApJ...928..156K},
	doi = {10.3847/1538-4357/ac5899},
	urldate = {2025-10-28},
	journal = {The Astrophysical Journal},
	author = {Krapp, Leonardo and Kratter, Kaitlin M. and Youdin, Andrew N.},
	month = apr,
	year = {2022},
	note = {Publisher: IOP
ADS Bibcode: 2022ApJ...928..156K},
	pages = {156},
}

@article{Li2025,
	title = {Discovery of {Hα} {Emission} from a {Protoplanet} {Candidate} around the {Young} {Star} {2MASS} {J16120668}–3010270 with {MagAO}-{X}},
	volume = {990},
	issn = {0004-637X},
	url = {https://ui.adsabs.harvard.edu/abs/2025ApJ...990L..70L},
	doi = {10.3847/2041-8213/adfcbd},
	urldate = {2025-10-28},
	journal = {The Astrophysical Journal},
	author = {Li, Jialin and Close, Laird M. and Long, Feng and Males, Jared R. and Haffert, Sebastiaan Y. and Weinberger, Alycia and Follette, Katherine and Andrews, Sean and Carpenter, John and Foster, Warren B. and Van Gorkom, Kyle and Hedglen, Alexander D. and Herczeg, Gregory J. and Johnson, Parker T. and Kautz, Maggie Y. and Kueny, Jay K. and Li, Rixin and Liberman, Joshua and Long, Joseph D. and Lumbres, Jennifer and Marino, Sebastian and Matrà, Luca and McEwen, Eden A. and Guyon, Olivier and Pearce, Logan A. and Pérez, Laura M. and Pinilla, Paola and Schatz, Lauren and Shi, Yangfan and Twitchell, Katie and Wagner, Kevin and Wilner, David and Wu, Ya-Lin and Zhang, Shangjia and Zhu, Zhaohuan},
	month = sep,
	year = {2025},
	note = {Publisher: IOP
ADS Bibcode: 2025ApJ...990L..70L},
	pages = {L70},
}

@article{Zhu2015,
	title = {{ACCRETING} {CIRCUMPLANETARY} {DISKS}: {OBSERVATIONAL} {SIGNATURES}},
	volume = {799},
	issn = {0004-637X},
	shorttitle = {{ACCRETING} {CIRCUMPLANETARY} {DISKS}},
	url = {https://dx.doi.org/10.1088/0004-637X/799/1/16},
	doi = {10.1088/0004-637X/799/1/16},
	language = {en},
	number = {1},
	urldate = {2024-12-04},
	journal = {The Astrophysical Journal},
	author = {Zhu, Zhaohuan},
	month = jan,
	year = {2015},
	note = {Publisher: The American Astronomical Society},
	pages = {16},
}

@article{Zhu2018,
	title = {On the radio detectability of circumplanetary discs},
	volume = {479},
	issn = {0035-8711},
	url = {https://doi.org/10.1093/mnras/sty1503},
	doi = {10.1093/mnras/sty1503},
	number = {2},
	urldate = {2024-12-04},
	journal = {Monthly Notices of the Royal Astronomical Society},
	author = {Zhu, Zhaohuan and Andrews, Sean M and Isella, Andrea},
	month = sep,
	year = {2018},
	pages = {1850--1865},
}

@article{Zhu2016,
	title = {{SHOCK}-{DRIVEN} {ACCRETION} {IN} {CIRCUMPLANETARY} {DISKS}: {OBSERVABLES} {AND} {SA}℡{LITE} {FORMATION}},
	volume = {832},
	issn = {0004-637X},
	shorttitle = {{SHOCK}-{DRIVEN} {ACCRETION} {IN} {CIRCUMPLANETARY} {DISKS}},
	url = {https://dx.doi.org/10.3847/0004-637X/832/2/193},
	doi = {10.3847/0004-637X/832/2/193},
	language = {en},
	number = {2},
	urldate = {2024-12-04},
	journal = {The Astrophysical Journal},
	author = {Zhu, Zhaohuan and Ju, Wenhua and Stone, James M.},
	month = dec,
	year = {2016},
	note = {Publisher: The American Astronomical Society},
	pages = {193},
}

@article{Adams2021,
	title = {A {Theoretical} {Framework} for the {Mass} {Distribution} of {Gas} {Giant} {Planets} {Forming} through the {Core} {Accretion} {Paradigm}},
	volume = {909},
	issn = {0004-637X},
	url = {https://ui.adsabs.harvard.edu/abs/2021ApJ...909....1A},
	doi = {10.3847/1538-4357/abdd2b},
	urldate = {2024-12-04},
	journal = {The Astrophysical Journal},
	author = {Adams, Fred C. and Meyer, Michael R. and Adams, Arthur D.},
	month = mar,
	year = {2021},
	note = {Publisher: IOP
ADS Bibcode: 2021ApJ...909....1A},
	pages = {1},
}

@article{Toomre1964,
	title = {On the gravitational stability of a disk of stars.},
	volume = {139},
	issn = {0004-637X},
	url = {https://ui.adsabs.harvard.edu/abs/1964ApJ...139.1217T},
	doi = {10.1086/147861},
	urldate = {2024-12-05},
	journal = {The Astrophysical Journal},
	author = {Toomre, A.},
	month = may,
	year = {1964},
	note = {Publisher: IOP
ADS Bibcode: 1964ApJ...139.1217T},
	pages = {1217--1238},
}

@article{Adams2022,
	title = {Analytic {Approach} to the {Late} {Stages} of {Giant} {Planet} {Formation}},
	volume = {934},
	issn = {0004-637X},
	url = {https://dx.doi.org/10.3847/1538-4357/ac7a3e},
	doi = {10.3847/1538-4357/ac7a3e},
	language = {en},
	number = {2},
	urldate = {2024-12-05},
	journal = {The Astrophysical Journal},
	author = {Adams, Fred C. and Batygin, Konstantin},
	month = jul,
	year = {2022},
	note = {Publisher: The American Astronomical Society},
	pages = {111},
}

@article{DAlessio1998,
	title = {Accretion {Disks} around {Young} {Objects}. {I}. {The} {Detailed} {Vertical} {Structure}},
	volume = {500},
	issn = {0004-637X},
	url = {https://ui.adsabs.harvard.edu/abs/1998ApJ...500..411D},
	doi = {10.1086/305702},
	urldate = {2024-12-17},
	journal = {The Astrophysical Journal},
	author = {D'Alessio, Paola and Cantö, Jorge and Calvet, Nuria and Lizano, Susana},
	month = jun,
	year = {1998},
	note = {Publisher: IOP
ADS Bibcode: 1998ApJ...500..411D},
	pages = {411--427},
}

@article{DAlessio2006,
	title = {Effects of {Dust} {Growth} and {Settling} in {T} {Tauri} {Disks}},
	volume = {638},
	issn = {0004-637X},
	url = {https://ui.adsabs.harvard.edu/abs/2006ApJ...638..314D},
	doi = {10.1086/498861},
	urldate = {2024-12-17},
	journal = {The Astrophysical Journal},
	author = {D'Alessio, Paola and Calvet, Nuria and Hartmann, Lee and Franco-Hernández, Ramiro and Servín, Hermelinda},
	month = feb,
	year = {2006},
	note = {Publisher: IOP
ADS Bibcode: 2006ApJ...638..314D},
	pages = {314--335},
}

@article{DAlessio1999,
	title = {Accretion {Disks} around {Young} {Objects}. {II}. {Tests} of {Well}-mixed {Models} with {ISM} {Dust}},
	volume = {527},
	issn = {0004-637X},
	url = {https://ui.adsabs.harvard.edu/abs/1999ApJ...527..893D},
	doi = {10.1086/308103},
	urldate = {2024-12-17},
	journal = {The Astrophysical Journal},
	author = {D'Alessio, Paola and Calvet, Nuria and Hartmann, Lee and Lizano, Susana and Cantó, Jorge},
	month = dec,
	year = {1999},
	note = {Publisher: IOP
ADS Bibcode: 1999ApJ...527..893D},
	pages = {893--909},
}

@article{DAlessio2001,
	title = {Accretion {Disks} around {Young} {Objects}. {III}. {Grain} {Growth}},
	volume = {553},
	issn = {0004-637X},
	url = {https://ui.adsabs.harvard.edu/abs/2001ApJ...553..321D},
	doi = {10.1086/320655},
	urldate = {2024-12-17},
	journal = {The Astrophysical Journal},
	author = {D'Alessio, Paola and Calvet, Nuria and Hartmann, Lee},
	month = may,
	year = {2001},
	note = {Publisher: IOP
ADS Bibcode: 2001ApJ...553..321D},
	pages = {321--334},
}

@article{Pollack1996,
	title = {Formation of the {Giant} {Planets} by {Concurrent} {Accretion} of {Solids} and {Gas}},
	volume = {124},
	issn = {0019-1035},
	url = {https://ui.adsabs.harvard.edu/abs/1996Icar..124...62P},
	doi = {10.1006/icar.1996.0190},
	urldate = {2024-12-17},
	journal = {Icarus},
	author = {Pollack, James B. and Hubickyj, Olenka and Bodenheimer, Peter and Lissauer, Jack J. and Podolak, Morris and Greenzweig, Yuval},
	month = nov,
	year = {1996},
	note = {ADS Bibcode: 1996Icar..124...62P},
	pages = {62--85},
}

@article{Cassen1981,
	title = {On the formation of protostellar disks},
	volume = {48},
	issn = {0019-1035},
	url = {https://ui.adsabs.harvard.edu/abs/1981Icar...48..353C/abstract},
	doi = {10.1016/0019-1035(81)90051-8},
	language = {en},
	number = {3},
	urldate = {2024-12-21},
	journal = {Icarus},
	author = {Cassen, P. and Moosman, A.},
	month = dec,
	year = {1981},
	pages = {353--376},
}

@article{Close2025,
	title = {Three {Years} of {High}-contrast {Imaging} of the {PDS} 70 b and c {Exoplanets} at {Hα} with {MagAO}-{X}: {Evidence} of {Strong} {Protoplanet} {Hα} {Variability} and {Circumplanetary} {Dust}},
	volume = {169},
	issn = {0004-6256},
	shorttitle = {Three {Years} of {High}-contrast {Imaging} of the {PDS} 70 b and c {Exoplanets} at {Hα} with {MagAO}-{X}},
	url = {https://ui.adsabs.harvard.edu/abs/2025AJ....169...35C},
	doi = {10.3847/1538-3881/ad8648},
	urldate = {2025-02-24},
	journal = {The Astronomical Journal},
	author = {Close, Laird M. and Males, Jared R. and Li, Jialin and Haffert, Sebastiaan Y. and Long, Joseph D. and Hedglen, Alexander D. and Weinberger, Alycia J. and Follette, Katherine B. and Apai, Daniel and Doyon, Rene and Foster, Warren and Gasho, Victor and Van Gorkom, Kyle and Guyon, Olivier and Kautz, Maggie Y. and Kueny, Jay and Lumbres, Jennifer and McLeod, Avalon and McEwen, Eden and Pavao, Clarissa and Pearce, Logan and Perez, Laura and Schatz, Lauren and Szulágyi, Judit and Wagner, Kevin and Wu, Ya-Lin},
	month = jan,
	year = {2025},
	note = {Publisher: IOP
ADS Bibcode: 2025AJ....169...35C},
	pages = {35},
}

@article{Viswanath2024,
	title = {{ExoplaNeT} {accRetion} {mOnitoring} {sPectroscopic} {surveY} ({ENTROPY}): {I}. {Evidence} for magnetospheric accretion in the young isolated planetary-mass object {2MASS} {J11151597}+1937266},
	volume = {691},
	issn = {0004-6361},
	shorttitle = {{ExoplaNeT} {accRetion} {mOnitoring} {sPectroscopic} {surveY} ({ENTROPY})},
	url = {https://ui.adsabs.harvard.edu/abs/2024A&A...691A..64V},
	doi = {10.1051/0004-6361/202450881},
	urldate = {2025-03-03},
	journal = {Astronomy and Astrophysics},
	author = {Viswanath, Gayathri and Ringqvist, Simon C. and Demars, Dorian and Janson, Markus and Bonnefoy, Mickaël and Aoyama, Yuhiko and Marleau, Gabriel-Dominique and Dougados, Catherine and Szulágyi, Judit and Thanathibodee, Thanawuth},
	month = nov,
	year = {2024},
	note = {Publisher: EDP
ADS Bibcode: 2024A\&A...691A..64V},
	pages = {A64},
}

@article{Thanathibodee2019,
	title = {Magnetospheric {Accretion} as a {Source} of {Hα} {Emission} from {Protoplanets} around {PDS} 70},
	volume = {885},
	issn = {0004-637X},
	url = {https://ui.adsabs.harvard.edu/abs/2019ApJ...885...94T},
	doi = {10.3847/1538-4357/ab44c1},
	urldate = {2025-03-03},
	journal = {The Astrophysical Journal},
	author = {Thanathibodee, Thanawuth and Calvet, Nuria and Bae, Jaehan and Muzerolle, James and Hernández, Ramiro Franco},
	month = nov,
	year = {2019},
	note = {Publisher: IOP
ADS Bibcode: 2019ApJ...885...94T},
	pages = {94},
}

@article{Benisty2021,
	title = {A {Circumplanetary} {Disk} around {PDS70c}},
	volume = {916},
	issn = {0004-637X},
	url = {https://ui.adsabs.harvard.edu/abs/2021ApJ...916L...2B},
	doi = {10.3847/2041-8213/ac0f83},
	urldate = {2025-03-03},
	journal = {The Astrophysical Journal},
	author = {Benisty, Myriam and Bae, Jaehan and Facchini, Stefano and Keppler, Miriam and Teague, Richard and Isella, Andrea and Kurtovic, Nicolas T. and Pérez, Laura M. and Sierra, Anibal and Andrews, Sean M. and Carpenter, John and Czekala, Ian and Dominik, Carsten and Henning, Thomas and Menard, Francois and Pinilla, Paola and Zurlo, Alice},
	month = jul,
	year = {2021},
	note = {Publisher: IOP
ADS Bibcode: 2021ApJ...916L...2B},
	pages = {L2},
}

@article{Ringqvist2023,
	title = {Resolved near-{UV} hydrogen emission lines at 40-{Myr} super-{Jovian} protoplanet {Delorme} 1 ({AB})b. {Indications} of magnetospheric accretion},
	volume = {669},
	issn = {0004-6361},
	url = {https://ui.adsabs.harvard.edu/abs/2023A&A...669L..12R},
	doi = {10.1051/0004-6361/202245424},
	urldate = {2025-03-03},
	journal = {Astronomy and Astrophysics},
	author = {Ringqvist, Simon C. and Viswanath, Gayathri and Aoyama, Yuhiko and Janson, Markus and Marleau, Gabriel-Dominique and Brandeker, Alexis},
	month = jan,
	year = {2023},
	note = {ADS Bibcode: 2023A\&A...669L..12R},
	pages = {L12},
}

@article{Isella2019,
	title = {Detection of {Continuum} {Submillimeter} {Emission} {Associated} with {Candidate} {Protoplanets}},
	volume = {879},
	issn = {0004-637X},
	url = {https://ui.adsabs.harvard.edu/abs/2019ApJ...879L..25I},
	doi = {10.3847/2041-8213/ab2a12},
	urldate = {2025-03-03},
	journal = {The Astrophysical Journal},
	author = {Isella, Andrea and Benisty, Myriam and Teague, Richard and Bae, Jaehan and Keppler, Miriam and Facchini, Stefano and Pérez, Laura},
	month = jul,
	year = {2019},
	note = {Publisher: IOP
ADS Bibcode: 2019ApJ...879L..25I},
	pages = {L25},
}

@article{Haffert2019,
	title = {Two accreting protoplanets around the young star {PDS} 70},
	volume = {3},
	issn = {2397-3366},
	url = {https://ui.adsabs.harvard.edu/abs/2019NatAs...3..749H},
	doi = {10.1038/s41550-019-0780-5},
	urldate = {2025-03-03},
	journal = {Nature Astronomy},
	author = {Haffert, S. Y. and Bohn, A. J. and de Boer, J. and Snellen, I. A. G. and Brinchmann, J. and Girard, J. H. and Keller, C. U. and Bacon, R.},
	month = jun,
	year = {2019},
	note = {ADS Bibcode: 2019NatAs...3..749H},
	pages = {749--754},
}

@article{Taylor2025,
	title = {Radiative signatures of circumplanetary disks and envelopes during the late stages of giant planet formation},
	volume = {425},
	issn = {0019-1035},
	url = {https://ui.adsabs.harvard.edu/abs/2025Icar..42516327T},
	doi = {10.1016/j.icarus.2024.116327},
	urldate = {2025-03-04},
	journal = {Icarus},
	author = {Taylor, Aster G. and Adams, Fred C.},
	month = jan,
	year = {2025},
	note = {Publisher: Elsevier
ADS Bibcode: 2025Icar..42516327T},
	pages = {116327},
}

@article{Taylor2024,
	title = {Formation and structure of circumplanetary disks and envelopes during the late stages of giant planet formation},
	volume = {415},
	issn = {0019-1035},
	url = {https://ui.adsabs.harvard.edu/abs/2024Icar..41516044T},
	doi = {10.1016/j.icarus.2024.116044},
	urldate = {2025-03-04},
	journal = {Icarus},
	author = {Taylor, Aster G. and Adams, Fred C.},
	month = jun,
	year = {2024},
	note = {Publisher: Elsevier
ADS Bibcode: 2024Icar..41516044T},
	pages = {116044},
}

@article{Zhou2025,
	title = {Evidence for {Variable} {Accretion} onto {PDS} 70 c and {Implications} for {Protoplanet} {Detections}},
	volume = {980},
	issn = {0004-637X},
	url = {https://ui.adsabs.harvard.edu/abs/2025ApJ...980L..39Z},
	doi = {10.3847/2041-8213/adb134},
	urldate = {2025-03-19},
	journal = {The Astrophysical Journal},
	author = {Zhou, Yifan and Bowler, Brendan P. and Sanghi, Aniket and Marleau, Gabriel-Dominique and Takasao, Shinsuke and Aoyama, Yuhiko and Hasegawa, Yasuhiro and Thanathibodee, Thanawuth and Uyama, Taichi and Hashimoto, Jun and Wagner, Kevin and Calvet, Nuria and Demars, Dorian and Wu, Ya-Lin and Biddle, Lauren I. and Haffert, Sebastiaan Y. and Bryan, Marta L.},
	month = feb,
	year = {2025},
	note = {Publisher: IOP
ADS Bibcode: 2025ApJ...980L..39Z},
	pages = {L39},
}

@article{Krapp2024,
	title = {A {Thermodynamic} {Criterion} for the {Formation} of {Circumplanetary} {Disks}},
	volume = {973},
	issn = {0004-637X},
	url = {https://ui.adsabs.harvard.edu/abs/2024ApJ...973..153K},
	doi = {10.3847/1538-4357/ad644a},
	urldate = {2025-04-07},
	journal = {The Astrophysical Journal},
	author = {Krapp, Leonardo and Kratter, Kaitlin M. and Youdin, Andrew N. and Benítez-Llambay, Pablo and Masset, Frédéric and Armitage, Philip J.},
	month = oct,
	year = {2024},
	note = {Publisher: IOP
ADS Bibcode: 2024ApJ...973..153K},
	pages = {153},
}

@article{Ayliffe2009,
	title = {Circumplanetary disc properties obtained from radiation hydrodynamical simulations of gas accretion by protoplanets},
	volume = {397},
	issn = {0035-8711},
	url = {https://doi.org/10.1111/j.1365-2966.2009.15002.x},
	doi = {10.1111/j.1365-2966.2009.15002.x},
	number = {2},
	urldate = {2025-04-07},
	journal = {Monthly Notices of the Royal Astronomical Society},
	author = {Ayliffe, Ben A. and Bate, Matthew R.},
	month = aug,
	year = {2009},
	pages = {657--665},
}

@article{Julia,
	title = {Julia: {A} {Fresh} {Approach} to {Numerical} {Computing}},
	volume = {59},
	issn = {0036-1445},
	shorttitle = {Julia},
	url = {https://epubs.siam.org/doi/10.1137/141000671},
	doi = {10.1137/141000671},
	number = {1},
	urldate = {2025-04-07},
	journal = {SIAM Review},
	author = {Bezanson, Jeff and Edelman, Alan and Karpinski, Stefan and Shah, Viral B.},
	month = jan,
	year = {2017},
	note = {Publisher: Society for Industrial and Applied Mathematics},
	pages = {65--98},
}

@article{Choksi2025,
	title = {Spectral energy distributions of disc-embedded accreting protoplanets},
	volume = {537},
	issn = {0035-8711},
	url = {https://ui.adsabs.harvard.edu/abs/2025MNRAS.537.2945C},
	doi = {10.1093/mnras/stae2530},
	urldate = {2025-04-07},
	journal = {Monthly Notices of the Royal Astronomical Society},
	author = {Choksi, Nick and Chiang, Eugene},
	month = mar,
	year = {2025},
	note = {Publisher: OUP
ADS Bibcode: 2025MNRAS.537.2945C},
	pages = {2945--2960},
}

@article{Makie,
	title = {Makie.jl: {Flexible} high-performance data visualization for {Julia}},
	volume = {6},
	issn = {2475-9066},
	shorttitle = {Makie.jl},
	url = {https://joss.theoj.org/papers/10.21105/joss.03349},
	doi = {10.21105/joss.03349},
	language = {en},
	number = {65},
	urldate = {2025-04-07},
	journal = {Journal of Open Source Software},
	author = {Danisch, Simon and Krumbiegel, Julius},
	month = sep,
	year = {2021},
	pages = {3349},
}

@article{Lambrechts2017,
	title = {Reduced gas accretion on super-{Earths} and ice giants},
	volume = {606},
	copyright = {© ESO, 2017},
	issn = {0004-6361, 1432-0746},
	url = {https://www.aanda.org/articles/aa/abs/2017/10/aa31014-17/aa31014-17.html},
	doi = {10.1051/0004-6361/201731014},
	language = {en},
	urldate = {2025-04-07},
	journal = {Astronomy \& Astrophysics},
	author = {Lambrechts, M. and Lega, E.},
	month = oct,
	year = {2017},
	note = {Publisher: EDP Sciences},
	pages = {A146},
}

@article{Lambrechts2019,
	title = {Quasi-static contraction during runaway gas accretion onto giant planets},
	volume = {630},
	copyright = {© ESO 2019},
	issn = {0004-6361, 1432-0746},
	url = {https://www.aanda.org/articles/aa/abs/2019/10/aa34413-18/aa34413-18.html},
	doi = {10.1051/0004-6361/201834413},
	language = {en},
	urldate = {2025-04-07},
	journal = {Astronomy \& Astrophysics},
	author = {Lambrechts, M. and Lega, E. and Nelson, R. P. and Crida, A. and Morbidelli, A.},
	month = oct,
	year = {2019},
	note = {Publisher: EDP Sciences},
	pages = {A82},
}

@article{Szulagyi2016,
	title = {Circumplanetary disc or circumplanetary envelope?},
	volume = {460},
	issn = {0035-8711},
	url = {https://doi.org/10.1093/mnras/stw1160},
	doi = {10.1093/mnras/stw1160},
	number = {3},
	urldate = {2025-04-07},
	journal = {Monthly Notices of the Royal Astronomical Society},
	author = {Szulágyi, J. and Masset, F. and Lega, E. and Crida, A. and Morbidelli, A. and Guillot, T.},
	month = aug,
	year = {2016},
	pages = {2853--2861},
}

@article{Dullemond2012,
	title = {{RADMC}-{3D}: {A} multi-purpose radiative transfer tool},
	shorttitle = {{RADMC}-{3D}},
	url = {https://ui.adsabs.harvard.edu/abs/2012ascl.soft02015D},
	urldate = {2025-04-07},
	journal = {Astrophysics Source Code Library},
	author = {Dullemond, C. P. and Juhasz, A. and Pohl, A. and Sereshti, F. and Shetty, R. and Peters, T. and Commercon, B. and Flock, M.},
	month = feb,
	year = {2012},
	note = {ADS Bibcode: 2012ascl.soft02015D},
	pages = {ascl:1202.015},
}

@article{Kawazoe1993,
	title = {Unstable {Accretion} {Disks} in {FU} {Orionis} {Stars}},
	volume = {45},
	issn = {0004-6264},
	url = {https://ui.adsabs.harvard.edu/abs/1993PASJ...45..715K},
	urldate = {2025-04-21},
	journal = {Publications of the Astronomical Society of Japan},
	author = {Kawazoe, Eiko and Mineshige, Shin},
	month = oct,
	year = {1993},
	note = {Publisher: OUP
ADS Bibcode: 1993PASJ...45..715K},
	pages = {715--725},
}

@article{Bell1994,
	title = {Using {FU} {Orionis} {Outbursts} to {Constrain} {Self}-regulated {Protostellar} {Disk} {Models}},
	volume = {427},
	issn = {0004-637X},
	url = {https://ui.adsabs.harvard.edu/abs/1994ApJ...427..987B},
	doi = {10.1086/174206},
	urldate = {2025-04-21},
	journal = {The Astrophysical Journal},
	author = {Bell, K. R. and Lin, D. N. C.},
	month = jun,
	year = {1994},
	note = {Publisher: IOP
ADS Bibcode: 1994ApJ...427..987B},
	pages = {987},
}

@book{Frank2002,
	title = {Accretion {Power} in {Astrophysics}: {Third} {Edition}},
	shorttitle = {Accretion {Power} in {Astrophysics}},
	url = {https://ui.adsabs.harvard.edu/abs/2002apa..book.....F},
	urldate = {2025-04-22},
	publisher = {Cambridge University Press},
	author = {Frank, Juhan and King, Andrew and Raine, Derek J.},
	month = jan,
	year = {2002},
	note = {Publication Title: Accretion Power in Astrophysics
ADS Bibcode: 2002apa..book.....F},
}

@book{Hartmann2009,
	title = {Accretion {Processes} in {Star} {Formation}: {Second} {Edition}},
	shorttitle = {Accretion {Processes} in {Star} {Formation}},
	url = {https://ui.adsabs.harvard.edu/abs/2009apsf.book.....H},
	urldate = {2025-04-22},
	publisher = {Cambridge University Press},
	author = {Hartmann, Lee},
	month = jan,
	year = {2009},
	note = {Publication Title: Accretion Processes in Star Formation: Second Edition
ADS Bibcode: 2009apsf.book.....H},
}

@article{Marley2021,
	title = {The {Sonora} {Brown} {Dwarf} {Atmosphere} and {Evolution} {Models}. {I}. {Model} {Description} and {Application} to {Cloudless} {Atmospheres} in {Rainout} {Chemical} {Equilibrium}},
	volume = {920},
	issn = {0004-637X},
	url = {https://ui.adsabs.harvard.edu/abs/2021ApJ...920...85M},
	doi = {10.3847/1538-4357/ac141d},
	urldate = {2025-04-23},
	journal = {The Astrophysical Journal},
	author = {Marley, Mark S. and Saumon, Didier and Visscher, Channon and Lupu, Roxana and Freedman, Richard and Morley, Caroline and Fortney, Jonathan J. and Seay, Christopher and Smith, Adam J. R. W. and Teal, D. J. and Wang, Ruoyan},
	month = oct,
	year = {2021},
	note = {Publisher: IOP
ADS Bibcode: 2021ApJ...920...85M},
	pages = {85},
}

@article{Shakura1973,
	title = {Black holes in binary systems. {Observational} appearance.},
	volume = {24},
	issn = {0004-6361},
	url = {https://ui.adsabs.harvard.edu/abs/1973A&A....24..337S},
	urldate = {2025-04-28},
	journal = {Astronomy and Astrophysics},
	author = {Shakura, N. I. and Sunyaev, R. A.},
	month = jan,
	year = {1973},
	note = {ADS Bibcode: 1973A\&A....24..337S},
	pages = {337--355},
}

@book{Armitage2020,
	title = {Astrophysics of planet formation, {Second} {Edition}},
	url = {https://ui.adsabs.harvard.edu/abs/2020apfs.book.....A},
	urldate = {2025-05-15},
	publisher = {Cambridge University Press},
	author = {Armitage, Philip J.},
	month = jan,
	year = {2020},
	note = {Publication Title: Astrophysics of planet formation
ADS Bibcode: 2020apfs.book.....A},
}

@article{Blandford1982,
	title = {Hydromagnetic flows from accretion disks and the production of radio jets.},
	volume = {199},
	issn = {0035-8711},
	url = {https://ui.adsabs.harvard.edu/abs/1982MNRAS.199..883B},
	doi = {10.1093/mnras/199.4.883},
	urldate = {2025-05-27},
	journal = {Monthly Notices of the Royal Astronomical Society},
	author = {Blandford, R. D. and Payne, D. G.},
	month = jun,
	year = {1982},
	note = {Publisher: OUP
ADS Bibcode: 1982MNRAS.199..883B},
	pages = {883--903},
}

@article{Ghosh1978,
	title = {Disk accretion by magnetic neutron stars.},
	volume = {223},
	issn = {0004-637X},
	url = {https://ui.adsabs.harvard.edu/abs/1978ApJ...223L..83G},
	doi = {10.1086/182734},
	urldate = {2025-05-27},
	journal = {The Astrophysical Journal},
	author = {Ghosh, P. and Lamb, F. K.},
	month = jul,
	year = {1978},
	note = {Publisher: IOP
ADS Bibcode: 1978ApJ...223L..83G},
	pages = {L83--L87},
}

@misc{Houge2025,
	title = {Burned to ashes: {How} the thermal decomposition of refractory organics in the inner protoplanetary disc impacts the gas-phase {C}/{O} ratio},
	shorttitle = {Burned to ashes},
	url = {https://ui.adsabs.harvard.edu/abs/2025arXiv250520427H},
	doi = {10.48550/arXiv.2505.20427},
	urldate = {2025-06-02},
	publisher = {arXiv},
	author = {Houge, Adrien and Johansen, Anders and Bergin, Edwin and Ciesla, Fred J. and Bitsch, Bertram and Lambrechts, Michiel and Henning, Thomas and Perotti, Giulia},
	month = may,
	year = {2025},
	note = {ADS Bibcode: 2025arXiv250520427H},
}

@inproceedings{Helled2014,
	address = {eprint: arXiv:1311.1142},
	title = {Giant {Planet} {Formation}, {Evolution}, and {Internal} {Structure}},
	url = {https://ui.adsabs.harvard.edu/abs/2014prpl.conf..643H},
	doi = {10.2458/azu_uapress_9780816531240-ch028},
	urldate = {2025-06-05},
	booktitle = {Protostars and {Planets} {VI}},
	author = {Helled, R. and Bodenheimer, P. and Podolak, M. and Boley, A. and Meru, F. and Nayakshin, S. and Fortney, J. J. and Mayer, L. and Alibert, Y. and Boss, A. P.},
	month = jan,
	year = {2014},
	note = {ADS Bibcode: 2014prpl.conf..643H},
	pages = {643--665},
}

@article{Pringle1981,
	title = {Accretion discs in astrophysics},
	volume = {19},
	issn = {0066-4146},
	url = {https://ui.adsabs.harvard.edu/abs/1981ARA&A..19..137P},
	doi = {10.1146/annurev.aa.19.090181.001033},
	urldate = {2025-06-14},
	journal = {Annual Review of Astronomy and Astrophysics},
	author = {Pringle, J. E.},
	month = jan,
	year = {1981},
	note = {ADS Bibcode: 1981ARA\&A..19..137P},
	pages = {137--162},
}

@article{Tobin2023,
	title = {Deuterium-enriched water ties planet-forming disks to comets and protostars},
	volume = {615},
	issn = {1476-4687},
	doi = {10.1038/s41586-022-05676-z},
	language = {eng},
	number = {7951},
	journal = {Nature},
	author = {Tobin, John J. and van 't Hoff, Merel L. R. and Leemker, Margot and van Dishoeck, Ewine F. and Paneque-Carreño, Teresa and Furuya, Kenji and Harsono, Daniel and Persson, Magnus V. and Cleeves, L. Ilsedore and Sheehan, Patrick D. and Cieza, Lucas},
	month = mar,
	year = {2023},
	pmid = {36890372},
	pages = {227--230},
}

@article{Harsono2015,
	title = {Volatile snowlines in embedded disks around low-mass protostars},
	volume = {582},
	issn = {0004-6361},
	url = {https://ui.adsabs.harvard.edu/abs/2015A&A...582A..41H},
	doi = {10.1051/0004-6361/201525966},
	urldate = {2025-06-16},
	journal = {Astronomy and Astrophysics},
	author = {Harsono, D. and Bruderer, S. and van Dishoeck, E. F.},
	month = oct,
	year = {2015},
	note = {ADS Bibcode: 2015A\&A...582A..41H},
	pages = {A41},
}

@article{Mayor1995,
	title = {A {Jupiter}-mass companion to a solar-type star},
	volume = {378},
	issn = {0028-0836},
	url = {https://ui.adsabs.harvard.edu/abs/1995Natur.378..355M},
	doi = {10.1038/378355a0},
	urldate = {2025-06-20},
	journal = {Nature},
	author = {Mayor, Michel and Queloz, Didier},
	month = nov,
	year = {1995},
	note = {ADS Bibcode: 1995Natur.378..355M},
	pages = {355--359},
}

@article{Aoyama2019,
	title = {Constraining {Planetary} {Gas} {Accretion} {Rate} from {Hα} {Line} {Width} and {Intensity}: {Case} of {PDS} 70 b and c},
	volume = {885},
	issn = {0004-637X},
	shorttitle = {Constraining {Planetary} {Gas} {Accretion} {Rate} from {Hα} {Line} {Width} and {Intensity}},
	url = {https://ui.adsabs.harvard.edu/abs/2019ApJ...885L..29A},
	doi = {10.3847/2041-8213/ab5062},
	urldate = {2025-06-23},
	journal = {The Astrophysical Journal},
	author = {Aoyama, Yuhiko and Ikoma, Masahiro},
	month = nov,
	year = {2019},
	note = {Publisher: IOP
ADS Bibcode: 2019ApJ...885L..29A},
	pages = {L29},
}

@article{Hashimoto2020,
	title = {Accretion {Properties} of {PDS} 70b with {MUSE}},
	volume = {159},
	issn = {0004-6256},
	url = {https://ui.adsabs.harvard.edu/abs/2020AJ....159..222H},
	doi = {10.3847/1538-3881/ab811e},
	urldate = {2025-06-23},
	journal = {The Astronomical Journal},
	author = {Hashimoto, Jun and Aoyama, Yuhiko and Konishi, Mihoko and Uyama, Taichi and Takasao, Shinsuke and Ikoma, Masahiro and Tanigawa, Takayuki},
	month = may,
	year = {2020},
	note = {Publisher: IOP
ADS Bibcode: 2020AJ....159..222H},
	pages = {222},
}

@article{Adams2025,
	title = {General {Analytic} {Solutions} for {Circumplanetary} {Disks} during the {Late} {Stages} of {Giant} {Planet} {Formation}},
	volume = {137},
	issn = {0004-6280},
	url = {https://ui.adsabs.harvard.edu/abs/2025PASP..137e4401A},
	doi = {10.1088/1538-3873/adcf57},
	urldate = {2025-06-23},
	journal = {Publications of the Astronomical Society of the Pacific},
	author = {Adams, Fred C. and Batygin, Konstantin},
	month = may,
	year = {2025},
	note = {Publisher: IOP
ADS Bibcode: 2025PASP..137e4401A},
	pages = {054401},
}

@article{Bowler2025,
	title = {Hα {Variability} of {AB} {Aur} b with the {Hubble} {Space} {Telescope}: {Probing} the {Nature} of a {Protoplanet} {Candidate} with {Accretion} {Light} {Echoes}},
	volume = {169},
	issn = {0004-6256},
	shorttitle = {Hα {Variability} of {AB} {Aur} b with the {Hubble} {Space} {Telescope}},
	url = {https://ui.adsabs.harvard.edu/abs/2025AJ....169..258B},
	doi = {10.3847/1538-3881/adb6a1},
	urldate = {2025-06-23},
	journal = {The Astronomical Journal},
	author = {Bowler, Brendan P. and Zhou, Yifan and Biddle, Lauren I. and Jiang, Lillian Yushu and Bae, Jaehan and Close, Laird M. and Follette, Katherine B. and Franson, Kyle and Kraus, Adam L. and Sanghi, Aniket and Tran, Quang and Ward-Duong, Kimberly and Wu, Ya-Lin and Zhu, Zhaohuan},
	month = may,
	year = {2025},
	note = {Publisher: IOP
ADS Bibcode: 2025AJ....169..258B},
	pages = {258},
}

@misc{Fasano2025,
	title = {Inner disc and circumplanetary material in the {PDS} 70 system},
	url = {https://ui.adsabs.harvard.edu/abs/2025arXiv250611709F},
	doi = {10.48550/arXiv.2506.11709},
	urldate = {2025-07-21},
	publisher = {arXiv},
	author = {Fasano, Daniele and Benisty, Myriam and Curone, Pietro and Facchini, Stefano and Zagaria, Francesco and Yoshida, Tomohiro C. and Doi, Kiyoaki and Sierra, Anibal and Andrews, Sean and Bae, Jaehan and Isella, Andrea and Kurtovic, Nicolás T. and Pérez, Laura M. and Pinilla, Paola and Rampinelli, Luna and Teague, Richard},
	month = jun,
	year = {2025},
	note = {ADS Bibcode: 2025arXiv250611709F},
}

@article{Thanathibodee2023,
	title = {A {Census} of the {Low} {Accretors}. {II}. {Accretion} {Properties}},
	volume = {944},
	issn = {0004-637X},
	url = {https://ui.adsabs.harvard.edu/abs/2023ApJ...944...90T},
	doi = {10.3847/1538-4357/acac84},
	urldate = {2025-08-11},
	journal = {The Astrophysical Journal},
	author = {Thanathibodee, Thanawuth and Molina, Brandon and Serna, Javier and Calvet, Nuria and Hernández, Jesús and Muzerolle, James and Franco-Hernández, Ramiro},
	month = feb,
	year = {2023},
	note = {Publisher: IOP
ADS Bibcode: 2023ApJ...944...90T},
	pages = {90},
}

@misc{Pittman2025,
	title = {The {ODYSSEUS} {Survey}. {Characterizing} magnetospheric geometries and hotspot structures in {T} {Tauri} stars},
	url = {https://ui.adsabs.harvard.edu/abs/2025arXiv250701162P},
	doi = {10.48550/arXiv.2507.01162},
	urldate = {2025-08-11},
	publisher = {arXiv},
	author = {Pittman, Caeley V. and Espaillat, C. C. and Robinson, Connor E. and Thanathibodee, Thanawuth and Lopez, Sophia and Calvet, Nuria and Zhu, Zhaohuan and Walter, Frederick M. and Wendeborn, John and Manara, Carlo Felice and Campbell-White, Justyn and Claes, Rik A. and Fang, Min and Frasca, Antonio and Gameiro, J. F. and Gangi, Manuele and Hernández, Jesus and Kóspál, Ágnes and Maucó, Karina and Muzerolle, James and Siwak, Michał and Tychoniec, Łukasz and Venuti, Laura},
	month = jul,
	year = {2025},
	note = {ADS Bibcode: 2025arXiv250701162P},
}

@article{Hartmann1996,
	title = {The {FU} {Orionis} {Phenomenon}},
	volume = {34},
	issn = {0066-4146},
	url = {https://ui.adsabs.harvard.edu/abs/1996ARA&A..34..207H},
	doi = {10.1146/annurev.astro.34.1.207},
	urldate = {2025-08-11},
	journal = {Annual Review of Astronomy and Astrophysics},
	author = {Hartmann, Lee and Kenyon, Scott J.},
	month = jan,
	year = {1996},
	note = {ADS Bibcode: 1996ARA\&A..34..207H},
	pages = {207--240},
}

@article{Ulrich1976,
	title = {An infall model for the {T} {Tauri} phenomenon.},
	volume = {210},
	issn = {0004-637X},
	url = {https://ui.adsabs.harvard.edu/abs/1976ApJ...210..377U},
	doi = {10.1086/154840},
	journal = {The Astrophysical Journal},
	author = {Ulrich, R. K.},
	month = dec,
	year = {1976},
	note = {tex.adsnote: Provided by the SAO/NASA Astrophysics Data System},
	pages = {377--391},
}

@article{Szulagyi2019,
	title = {Observability of forming planets and their circumplanetary discs {II}. - {SEDs} and near-infrared fluxes},
	volume = {487},
	issn = {0035-8711},
	url = {https://ui.adsabs.harvard.edu/abs/2019MNRAS.487.1248S},
	doi = {10.1093/mnras/stz1326},
	urldate = {2025-10-29},
	journal = {Monthly Notices of the Royal Astronomical Society},
	author = {Szulágyi, J. and Dullemond, C. P. and Pohl, A. and Quanz, S. P.},
	month = jul,
	year = {2019},
	note = {Publisher: OUP
ADS Bibcode: 2019MNRAS.487.1248S},
	pages = {1248--1258},
}

@article{Micolta2024,
	title = {Using the {Ca} {II} {Lines} in {T} {Tauri} {Stars} to {Infer} the {Abundance} of {Refractory} {Elements} in the {Innermost} {Disk} {Region}},
	volume = {976},
	issn = {0004-637X},
	url = {https://ui.adsabs.harvard.edu/abs/2024ApJ...976..251M},
	doi = {10.3847/1538-4357/ad8884},
	urldate = {2025-11-19},
	journal = {The Astrophysical Journal},
	author = {Micolta, Marbely and Calvet, Nuria and Thanathibodee, Thanawuth and Magris C., Gladis and Manara, Carlo F. and Venuti, Laura and Alcalá, Juan Manuel and Herczeg, Gregory J.},
	month = dec,
	year = {2024},
	note = {Publisher: IOP
ADS Bibcode: 2024ApJ...976..251M},
	pages = {251},
}

@article{Zhu2007,
	title = {The {Hot} {Inner} {Disk} of {FU} {Orionis}},
	volume = {669},
	issn = {0004-637X},
	url = {https://iopscience.iop.org/article/10.1086/521345/meta},
	doi = {10.1086/521345},
	language = {en},
	number = {1},
	urldate = {2025-11-10},
	journal = {The Astrophysical Journal},
	author = {Zhu, Zhaohuan and Hartmann, Lee and Calvet, Nuria and Hernandez, Jesus and Muzerolle, James and Tannirkulam, Ajay-Kumar},
	month = nov,
	year = {2007},
	note = {Publisher: IOP Publishing},
	pages = {483},
}

@article{Christiaens2024,
	title = {{MINDS}: {JWST}/{NIRCam} imaging of the protoplanetary disk {PDS} 70},
	journal = {arXiv e-prints},
	author = {Christiaens, V. and Samland, M. and Henning, Th. and Portilla-Revelo, B. and Perotti, G. and Matthews, E. and Absil, O. and Decin, L. and Kamp, I. and Boccaletti, A. and Tabone, B. and Marleau, G. -D. and van Dishoeck, E. F. and Güdel, M. and Lagage, P. -O. and Barrado, D. and Garatti, A. Caratti o and Glauser, A. M. and Olofsson, G. and Ray, T. P. and Scheithauer, S. and Vandenbussche, B. and Waters, L. B. F. M. and Arabhavi, A. M. and Grant, S. L. and Jang, H. and Kanwar, J. and Schreiber, J. and Schwarz, K. and Temmink, M. and Östlin, G.},
	month = mar,
	year = {2024},
	note = {arXiv: 2403.04855 [astro-ph.EP]
Number: arXiv:2403.04855
tex.adsnote: Provided by the SAO/NASA Astrophysics Data System},
	pages = {arXiv:2403.04855},
}

@article{Szulagyi2018,
	title = {Observability of forming planets and their circumplanetary discs – {I}. {Parameter} study for {ALMA}},
	volume = {473},
	issn = {0035-8711},
	url = {https://doi.org/10.1093/mnras/stx2602},
	doi = {10.1093/mnras/stx2602},
	number = {3},
	urldate = {2025-10-29},
	journal = {Monthly Notices of the Royal Astronomical Society},
	author = {Szulágyi, J. and Plas, G. van der and Meyer, M. R. and Pohl, A. and Quanz, S. P. and Mayer, L. and Daemgen, S. and Tamburello, V.},
	month = jan,
	year = {2018},
	pages = {3573--3583},
}

@article{Chen2022,
	title = {Observability of forming planets and their circumplanetary discs – {IV}. {With} {JWST} and {ELT}},
	volume = {516},
	issn = {0035-8711},
	url = {https://doi.org/10.1093/mnras/stac1976},
	doi = {10.1093/mnras/stac1976},
	number = {1},
	urldate = {2025-10-29},
	journal = {Monthly Notices of the Royal Astronomical Society},
	author = {Chen, Xueqing and Szulágyi, Judit},
	month = oct,
	year = {2022},
	pages = {506--528},
}

@article{Cugno2024,
	title = {Mid-infrared spectrum of the disk around the forming companion {GQ} lup {B} revealed by {JWST}/{MIRI}},
	doi = {10.48550/arXiv.2404.07086},
	journal = {arXiv e-prints},
	author = {Cugno, Gabriele and Patapis, Polychronis and Banzatti, Andrea and Meyer, Michael and Dannert, Felix A. and Stolker, Tomas and MacDonald, Ryan J. and Pontoppidan, Klaus M.},
	month = apr,
	year = {2024},
	note = {arXiv: 2404.07086 [astro-ph.EP]
Number: arXiv:2404.07086
tex.adsnote: Provided by the SAO/NASA Astrophysics Data System},
	pages = {arXiv:2404.07086},
}

@article{Chevalier1983,
	title = {The enviroments of {T} {Tauri} stars},
	volume = {268},
	issn = {0004-637X},
	url = {https://ui.adsabs.harvard.edu/abs/1983ApJ...268..753C},
	doi = {10.1086/160997},
	urldate = {2025-08-12},
	journal = {The Astrophysical Journal},
	author = {Chevalier, R. A.},
	month = may,
	year = {1983},
	note = {Publisher: IOP
ADS Bibcode: 1983ApJ...268..753C},
	pages = {753--765},
}

@article{Lubow1999,
	title = {Disk {Accretion} onto {High}-{Mass} {Planets}},
	volume = {526},
	issn = {0004-637X},
	url = {https://ui.adsabs.harvard.edu/abs/1999ApJ...526.1001L},
	doi = {10.1086/308045},
	urldate = {2025-08-12},
	journal = {The Astrophysical Journal},
	author = {Lubow, S. H. and Seibert, M. and Artymowicz, P.},
	month = dec,
	year = {1999},
	note = {Publisher: IOP
ADS Bibcode: 1999ApJ...526.1001L},
	pages = {1001--1012},
}

@article{Canup2002,
	title = {Formation of the {Galilean} {Satellites}: {Conditions} {ofAccretion}},
	volume = {124},
	issn = {1538-3881},
	shorttitle = {Formation of the {Galilean} {Satellites}},
	url = {https://iopscience.iop.org/article/10.1086/344684/meta},
	doi = {10.1086/344684},
	language = {en},
	number = {6},
	urldate = {2025-08-12},
	journal = {The Astronomical Journal},
	author = {Canup, Robin M. and Ward, William R.},
	month = dec,
	year = {2002},
	note = {Publisher: IOP Publishing},
	pages = {3404},
}

@misc{Dominguez-Jamett2025,
	title = {Multi-frequency observations of {PDS} 70c: {Radio} emission mechanisms in the circum-planetary environment},
	shorttitle = {Multi-frequency observations of {PDS} 70c},
	url = {https://ui.adsabs.harvard.edu/abs/2025arXiv250721970D},
	doi = {10.48550/arXiv.2507.21970},
	urldate = {2025-08-12},
	publisher = {arXiv},
	author = {Dominguez-Jamett, Oriana and Casassus, Simon and Liu, Hauyu Baobab and Aoyama, Yuhiko and Carcamo, Miguel and Weber, Philipp and Chrenko, Ondrej and Marleau, Gabriel-Dominique and Ercolano, Barbara and Szulagyi, Judit},
	month = jul,
	year = {2025},
	note = {ADS Bibcode: 2025arXiv250721970D},
}

@article{Martin2011,
	title = {Tidal truncation of circumplanetary discs},
	volume = {413},
	issn = {0035-8711},
	url = {https://doi.org/10.1111/j.1365-2966.2011.18228.x},
	doi = {10.1111/j.1365-2966.2011.18228.x},
	number = {2},
	urldate = {2025-08-12},
	journal = {Monthly Notices of the Royal Astronomical Society},
	author = {Martin, Rebecca G. and Lubow, Stephen H.},
	month = may,
	year = {2011},
	pages = {1447--1461},
}

@article{Quillen1998,
	title = {Do {Proto}-jovian {Planets} {Drive} {Outflows}?},
	volume = {508},
	issn = {0004-637X},
	url = {https://iopscience.iop.org/article/10.1086/306421/meta},
	doi = {10.1086/306421},
	language = {en},
	number = {2},
	urldate = {2025-08-13},
	journal = {The Astrophysical Journal},
	author = {Quillen, A. C. and Trilling, D. E.},
	month = dec,
	year = {1998},
	note = {Publisher: IOP Publishing},
	pages = {707},
}

@article{Mathis1977,
	title = {The size distribution of interstellar grains.},
	volume = {217},
	issn = {0004-637X},
	url = {https://ui.adsabs.harvard.edu/abs/1977ApJ...217..425M},
	doi = {10.1086/155591},
	urldate = {2025-08-29},
	journal = {The Astrophysical Journal},
	author = {Mathis, J. S. and Rumpl, W. and Nordsieck, K. H.},
	month = oct,
	year = {1977},
	note = {ADS Bibcode: 1977ApJ...217..425M},
	pages = {425--433},
}

@article{Close2025a,
	title = {Wide {Separation} {Planets} in {Time} ({WISPIT}): {Discovery} of a {Gap} {Hα} {Protoplanet} {WISPIT} 2b with {MagAO}-{X}},
	volume = {990},
	issn = {0004-637X},
	shorttitle = {Wide {Separation} {Planets} in {Time} ({WISPIT})},
	url = {https://ui.adsabs.harvard.edu/abs/2025ApJ...990L...9C},
	doi = {10.3847/2041-8213/adf7a5},
	urldate = {2025-09-02},
	journal = {The Astrophysical Journal},
	author = {Close, Laird M. and van Capelleveen, Richelle F. and Weible, Gabriel and Wagner, Kevin and Haffert, Sebastiaan Y. and Males, Jared R. and Ilyin, Ilya and Kenworthy, Matthew A. and Li, Jialin and Long, Joseph D. and Ertel, Steve and Ginski, Christian and Weinberger, Alycia J. and Follette, Kate and Liberman, Joshua and Twitchell, Katie and Johnson, Parker and Kueny, Jay and Apai, Daniel and Doyon, Rene and Foster, Warren and Gasho, Victor and Van Gorkom, Kyle and Guyon, Olivier and Kautz, Maggie Y. and McLeod, Avalon and McEwen, Eden and Pearce, Logan and Schatz, Lauren and Hedglen, Alexander D. and Wu, Ya-Lin and Isbell, Jacob and Power, Jenny and Carlson, Jared and Close, Emmeline and Tonucci, Elena and Mars, Matthijs},
	month = sep,
	year = {2025},
	note = {ADS Bibcode: 2025ApJ...990L...9C},
	pages = {L9},
}

@article{Calvet1998,
	title = {The {Structure} and {Emission} of the {Accretion} {Shock} in {T} {Tauri} {Stars}},
	volume = {509},
	issn = {0004-637X},
	url = {https://ui.adsabs.harvard.edu/abs/1998ApJ...509..802C},
	doi = {10.1086/306527},
	urldate = {2025-09-05},
	journal = {The Astrophysical Journal},
	author = {Calvet, Nuria and Gullbring, Erik},
	month = dec,
	year = {1998},
	note = {ADS Bibcode: 1998ApJ...509..802C},
	pages = {802--818},
}

\end{document}